%% file: paper.tex
\documentclass[conference]{IEEEtran}

\usepackage{tikz}
\usetikzlibrary{positioning}

\usepackage[outdir=./]{epstopdf}
\usepackage{algorithm}
\usepackage{algorithmic}
\usepackage{balance}

\usepackage{caption}
\usepackage{subcaption}
\usepackage{cite}

\newcommand\blfootnote[1]{%
  \begingroup
  \renewcommand\thefootnote{}\footnote{#1}%
  \addtocounter{footnote}{-1}%
  \endgroup
}

\begin{document}
\title{Using Application Data for SLA-aware\\Auto-scaling in Cloud Environments}

\author{\IEEEauthorblockN{Andre Abrantes D. P. Souza, Marco A. S. Netto}
\IEEEauthorblockA{IBM Research}
}
\maketitle
\blfootnote{The final publication is available at ieee.org/ieeexplore IEEE MASCOTS'15}

\begin{abstract}
%\boldmath
With the establishment of cloud computing as the environment of choice for most modern applications, auto-scaling is an economic matter of great importance. For applications like stream computing that process ever changing amounts of data, modifying the number and configuration of resources to meet performance requirements becomes essential. Current solutions on auto-scaling are mostly rule-based using infrastructure level metrics such as CPU/memory/network utilization, and system level metrics such as throughput and response time. In this paper, we introduce a study on how effective auto-scaling can be using data generated by the application itself. To make this assessment, two algorithms are proposed that use \textit{a priori} knowledge of the data stream and use sentiment analysis from soccer-related tweets, triggering auto-scaling operations according to rapid changes in the public sentiment about the soccer players that happens just before big bursts of messages. Our application-based auto-scaling was able to reduce the number of SLA violations by up to 95\% and reduce resource requirements by up to 33\%.

\end{abstract}

\IEEEpeerreviewmaketitle

\input{intro}

\input{autoscaling}

\input{fama}

\input{simulator}

%RESULTS NOT COMPILING
\input{results}

\input{conclusion}

\section*{Acknowledgment}

We thank Paulo Rodrigo Cavalin for his help with the sentiment analysis application. This work has been supported and partially funded by
FINEP / MCTI, under subcontract no. 03.14.0062.00.




%\bibliographystyle{IEEEtran}

\bibliographystyle{IEEEtran}

\balance

%\printbibliography
%\theendnotes

\bibliography{references}

\end{document}

%% file: intro.tex
\section{Introduction}

Cloud was initially created to host web applications but has become mature enough to host more complex applications, such as those in the big data space. Due to the large resource consumption from these new cloud applications, users are caution on how much they spend in the cloud to meet their QoS requirements. In this scenario, auto-scaling, also known as \textit{elasticity} \cite{hwang2015cloud}, is an important technique to help users configure resource allocation dynamically.

There is a large body of work in the literature about auto-scaling solutions 
\cite{aliedin2013workload,fadika2011delma,mao2011autoscaling,mao2010autoscaling,sedaghat2013virtual,shen2011cloudscale}. Most of the existing solutions are based on rules \cite{botran2014review} that assess system or infrastructure level variables. An example of a CPU-based threshold rule is: ``increase 10\% of resources if CPU usage is above 80\% for the last 5 minutes''. Other examples of auto-scaling metrics are memory, network, and storage usage, response time, and throughput.

Another source of metrics to trigger auto-scaling operations comes from the applications themselves. A signal inside the data generated by an application can serve as an earlier indicator that there will be a load change in near future. This signal can be more effective than waiting for CPU or network reach to undesirable utilization levels. Examples of signals are (i) a relevant news in a web site that was just published that may increase user access to the site; (ii) a data mining application that reaches an intermediate result that intensifies the use of computing power to explore more a search area; and (iii) a financial application that detects an unexpected trend that requires additional simulations to handle it.

In this paper, we carry out a study on using application data as a trigger for auto-scaling operations. Our hypothesis is that this approach meets QoS requirements more efficiently than using auto-scaling triggers based on infrastructure or system metrics. Therefore, our contributions are:

\begin{itemize}
\item Identification of auto-scaling triggers that use correlation between data produced by the application and the volume of data to be processed ($\S$ \ref{sec:app});
\item Two auto-scaling triggers based on application data, including one with user sentiment analysis ($\S$ \ref{sec:simulator});
\item An extensive evaluation of the auto-scaling triggers using millions of tweets from the 2013 FIFA Confederations Cup and an application that calculates public sentiment changes during soccer matches. We use a CPU-based threshold algorithm for comparison purposes ($\S$ \ref{sec:results}).
\end{itemize}

%% file: autoscaling.tex
\section{Background}

%Cloud services are mostly offered in one of the following models: Infrastructure-as-a-Service (IaaS), Platform-as-a-Service (PaaS) or Software-as-a-Service (SaaS). While SaaS applications rely on lower layers for auto-scaling by maintaining its functionality transparent to users, IaaS and PaaS must themselves deal with auto-scaling of resources according to their needs.

Auto-scaling is an important part of cloud computing as it serves to both keep up with a high resource utilization and save money when resources are underutilized. Manually managing resources is far inefficient for most applications as performance and input size usually vary over time. Automatically scaling applications is preferable because resources can be deployed faster and it can be done according to a great array of performance parameters beyond ordinary human capabilities.

The main auto-scaling operations are scale-in, scale-out, scale-up, and scale-down. Scale-in/out expands and shrinks the number of computing resources and scale-up/down expands and shrinks the computing power of existing resources. The first two operations are known as \textit{horizontal auto-scaling}, whereas the last two are known as \textit{vertical auto-scaling}. There are efforts in auto-scaling from both industry and academia.

%In PaaS environments, auto-scaling can also be horizontal or vertical. One could increase the number of instances of their jobs, thus replicating the service and increasing the system throughput, or could allocate more CPU (either by adding more processing cores or by increase each core capacity) or memory. In IaaS environments, however, a change in the configuration of the virtual machine is necessary, which leads to more challenging situations.

Amazon CloudWatch \cite{amazon} is a monitoring system to help users decide when cloud resources need to be modified. In this system, users specify upper and lower bounds for monitored metrics such as memory and CPU usage. Microsoft Azure Auto-scaling system \cite{microsoft} also allows users to specify these auto-scaling parameters. Scryer \cite{scryer}, from Netflix, is an auto-scaling engine that uses predictive models to know when resources should be added or removed. Its auto-scaling strategy is not exposed to users so they do not need to interact or specify auto-scaling thresholds and policies.

Ming et al.~\cite{mao2010autoscaling} proposed an architecture that deals with auto-scaling focusing on meeting user deadlines. Shen  {\it et al.} \cite{shen2011cloudscale} presented a system to automate elastic resource scaling for cloud computing environments. Their system does not require prior knowledge about the applications running in the cloud. Other projects consider auto-scaling in different scenarios, such as auto-scaling for MapReduce applications \cite{chen2012interactive,fadika2011delma}, vertical versus horizontal auto-scaling \cite{sedaghat2013virtual}, operational costs \cite{mao2011autoscaling}, and integer model based auto-scaling \cite{mao2010autoscaling}. Ali-Eldin  {\it et al.} \cite{aliedin2013workload} introduced a tool to analyze and classify workloads and assign the most suitable auto-scale controllers based on workload characteristics. Ali-Eldin  {\it et al.} also identified the challenge aspect of developing workload predictors. Cunha {\it et al.} \cite{renato2014patience} explored the use of user patience information to make better auto-scaling decisions. Netto {\it et al.} \cite{netto2014evaluating} introduced the concept of Auto-scaling Demand Index to determine how well auto-scaling operations are performing and presented a study to help users configure auto-scaling parameters. 

From all these works, it can be noticed that traditional auto-scaling techniques are similar for both PaaS and IaaS; \textit{i.e.} simple threshold-based rules using, for instance, CPU and memory as metrics to be monitored. Other parameters could be used for auto-scaling, for instance, application parameters. While in IaaS, the cloud infrastructure should only be aware of the OS level, in PaaS there is the possibility of the application being aware of the cloud infrastructure needs for auto-scaling. In order to simplify resource allocation decisions, Copil {\it et al.} \cite{copil2014advise} introduced a framework to advise on elasticity operations via time series analysis and Leitner {\it et al.} \cite{leitner2013data} explored application data and domain experts to avoid Service Level Agreement (SLA) violations.

Data stream applications, in particular, can be scaled in more than one dimension. They can be scaled by parallelizing operators or by increasing the quota of available resources for the application. But parallelization of operators does not tend to deliver a significant benefit to the user if the operator is CPU-bound since, most of the time, a single operator is capable of maximizing the usage of the available resources. There are also efforts on auto-scaling for data streaming applications \cite{kumbhare2013exploiting,heinze2014autoscaling,heinzeu2014latency}.

For data streaming applications, the resource management software can provide system related data such as input and output rates and queue sizes. This would most likely already improve auto-scaling systems. Trends in input rates could be found and output based SLA could be used. But there is still a third level of data that could be used for this kind of application: their own output.

The main novelty of this paper is to provide a real use case on how application data can be used for auto-scaling in practice and how beneficial this approach is compared to auto-scaling based on common infrastructure and system metrics.

%% file: fama.tex
\section{Use Case Application}
\label{sec:app}

We used an in-house application \cite{cavalin2014realtime,cavalin2015scalable} to study the impact of using application data as a trigger for auto-scaling. This application is based on IBM Streams and evaluates tweet sentiment at real time. The scenario explored here is in the context of analyzing public sentiment about players during soccer matches.

The application uses Twitter APIs to continuously read a live stream of tweets. To setup the reading stream, the application passes a set of keywords and a target language so that every tweet matching those criteria is sent to the client. Tweets come JSON-encoded and with a variety of data and meta-data, such as the author username and profile.

\begin{figure}[!t]
	\centering
	\begin{tikzpicture}
	[box/.style={draw,rounded corners,text width=3.3cm,align=center}]
	\node[box] (a) {(1) read and parse tweet};
	\node[box,below=of a] (b) {(2) check if it is about soccer};
	\node[box,below=of b] (c) {(3) get topics and sentiment};
	\node[box,below=of c] (d) {(4) extract terms};
	\node[box,right=of d] (e) {(5) accumulate statistics};
	\draw[->] (a) -- (b);\draw[->] (a) -| (e);
	\draw[->] (b) -- (c);\draw[->] (b) -| (e);
	\draw[->] (c) -- (d);\draw[->] (c) -| (e);
	\draw[->] (d) -- (e);
	\end{tikzpicture}
	\caption{Sentiment analysis application graph \cite{cavalin2014realtime,cavalin2015scalable}.}
	\label{fig:fama_pipeline}
\end{figure}
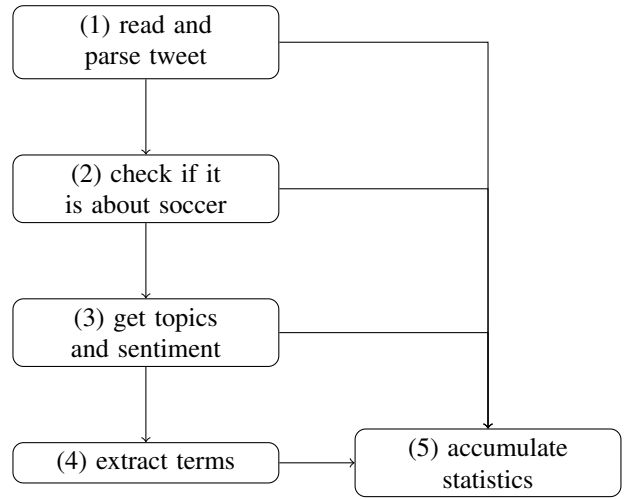

Figure \ref{fig:fama_pipeline} shows the application operator graph. Each block is a Processing Element (PE), \textit{i.e.} a set of operators abstracted to a higher level. Arrows represent the stream of data among the PEs and thus the different paths a tweet can take when traversing the graph. We define in the context of this paper that the path that the tweet takes through the graph defines its class.

Tweets that are completely used by the pipeline go through all PEs. However, most tweets are discarded in the processes, \textit{e.g.} a tweet might contain a particular keyword but actually have another subject than soccer. All discarded tweets are nevertheless sent to the final statistic accumulator node.

PEs (2), (3) and (4) in Figure \ref{fig:fama_pipeline} are actually very parallelized so they can better benefit from multiple CPUs and hosts. The source and sink PEs, (1) and (5), on the other hand, process one tweet at a time. But since their job is way simpler than that of the other PEs, they are not bottlenecks in the graph.

The way sentiment analysis application is implemented, the sentiment-related data is loaded once and the application does not need to consult databases or external APIs at runtime. Therefore, the application is not I/O-bound; the Tweeter API is its only possible I/O bottleneck and hence the application is mostly CPU-bound.

Since the application monitors live soccer matches, its infrastructure must support the variable and sometimes huge volume of tweets posted and deliver sentiment analysis at real time. That is a requisite from clients and the usual SLA agreed is that every tweet must be processed under 5 minutes.

\subsection{Sentiment analysis and tweet volume relationship}
\label{sec:sentiment_analysis}

For each tweet analyzed, sentiment is given as three real numbers called the probability that the tweet is positive, negative or neutral. These three numbers always sum to 1. The probability calculation is given by a machine learning based sentiment analysis, which is part of the in-house application \cite{cavalin2014realtime,cavalin2015scalable}.

To account for periods of high fluctuations in the sentiment time series, an exponential moving average is used. Using a window of one minute, a considerable correlation has been found between the sentiment at a given time and the number of tweets posted on the following minutes.

\begin{figure}
	\centering
	\includegraphics[width=0.9\linewidth]{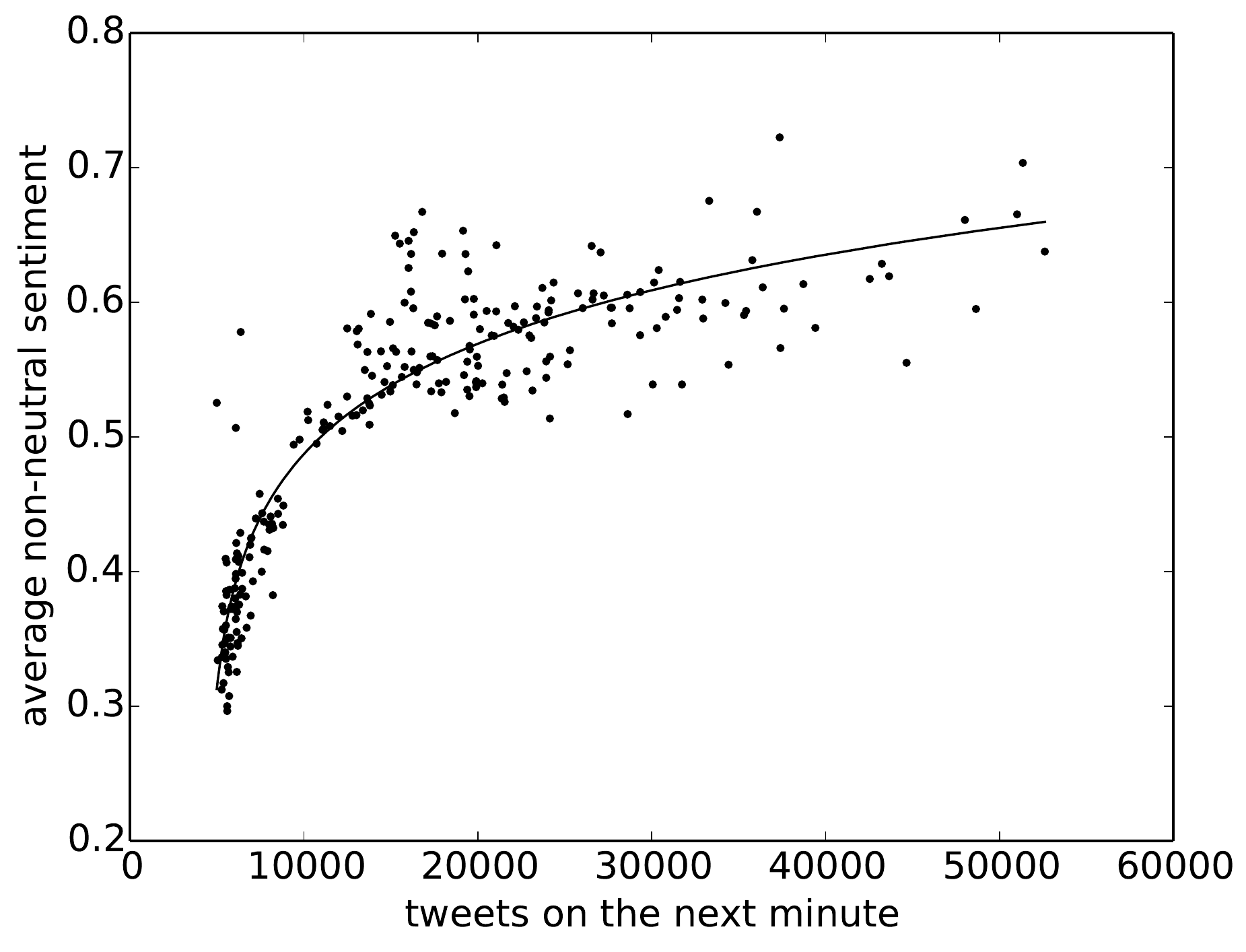}
	\caption{Relationship of the average sentiment on a given minute with the volume of tweets posted on the next minute for the Brazil vs Spain match.}
	\label{fig:sentiment_correlation}
\end{figure}

Figure \ref{fig:sentiment_correlation} shows the correlation between tweet sentiment and volume for the Brazil vs Spain match. There is a clear tendency that the more intense the sentiment the more tweets are posted. From the figure, it also seems that points are divided in two clusters. The first is a well behaved set of points with moderate sentiment, roughly below $0.4$. The second set, however, is spread on a broader area but has consistently higher tweet volumes.

Although the tweet volume is not easily predictable from the current sentiment level, there is a clear relationship between sentiment intensity and tweet volume in the following minutes. Table \ref{tab:sentiment_correlation} makes that correlation clear by showing the Pearson correlation coefficient for average sentiment level of a minute and the volume of tweets on the near future. Correlation of sentiment on time $t$ with the volume on time $t$ has the highest value of $0.79$ and decays slowly for the next 6 minutes before a significant variation.

\begin{table}
	\centering
	\caption{Sentiment correlation of the volume tweet at a given time with sentiment of time $t$.}
	\begin{tabular}{cc}
		\hline
		\hline
		\textbf{time} & \textbf{correlation} \\
		\hline
		 $t$ & 0.79 \\
		 $t+1$ & 0.78 \\
		$t+2$ & 0.76 \\
		$t+3$ & 0.76 \\
		$t+4$ & 0.76 \\
		$t+5$ & 0.75 \\
		$t+6$ & 0.75 \\
		$t+7$ & 0.74 \\
		$t+8$ & 0.72 \\
		$t+9$ & 0.71 \\
		$t+10$ & 0.70 \\
		\hline 
		\hline 
	\end{tabular}
	\label{tab:sentiment_correlation}
\end{table}

\begin{figure}
	\centering
	\includegraphics[width=0.9\linewidth]{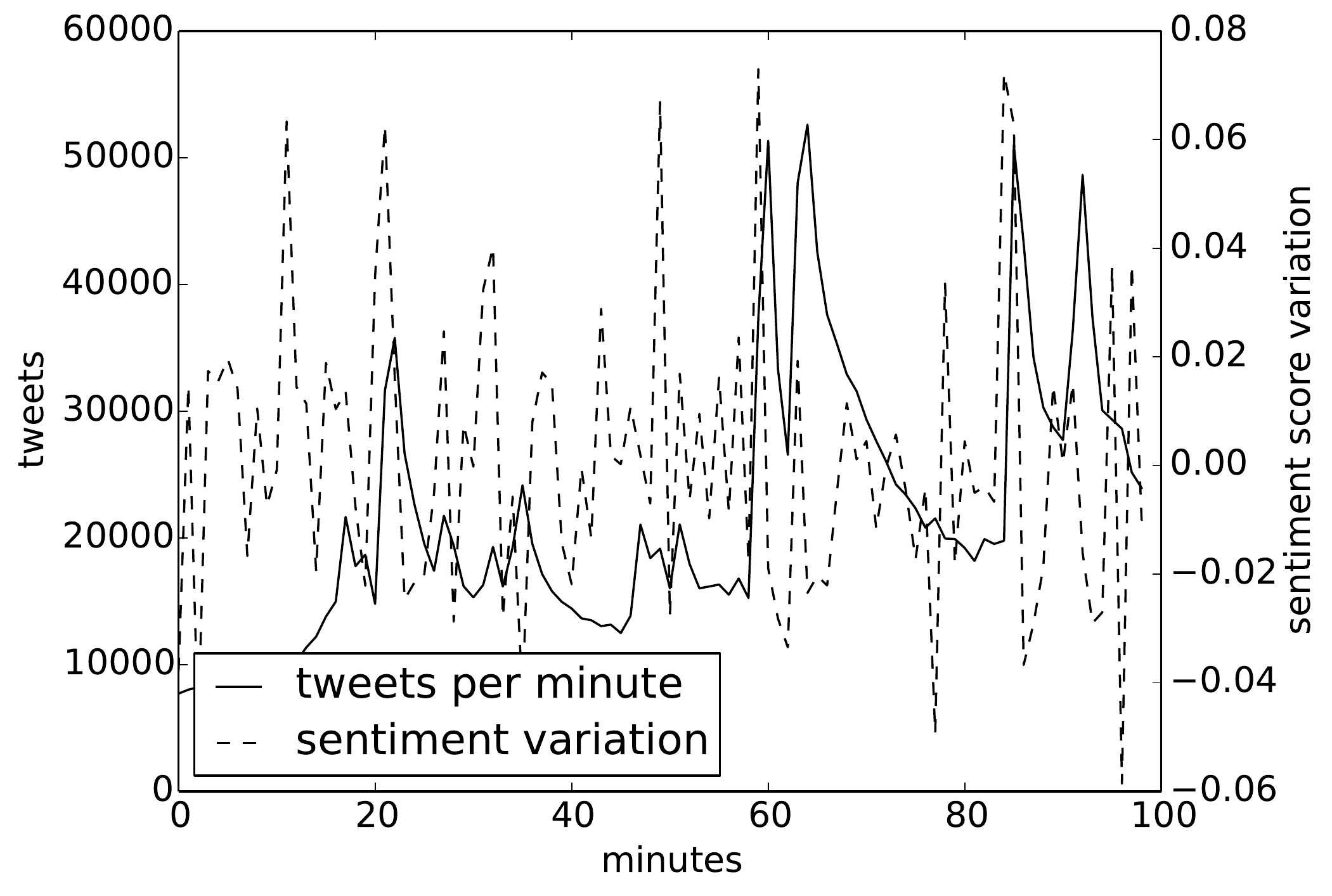}
	\caption{Sentiment variation and bursts of tweets.}
	\label{fig:sentiment_var}
\end{figure}

The sentiment is above 0.4 for most part of the matches and for most matches. This makes it hard to detect sudden burst of tweets simply by looking at the average \textit{sentiment score}\footnote{Sentiment score: tweet probability of being positive or negative.}. By analyzing the variation in sentiment time series we observe that bursts of tweets are preceded by a high sentiment variation. Figure \ref{fig:sentiment_var} shows how that happens over a period of 100 minutes of the Brazil vs Spain match. Although there are some false positives and a false negative in the example, peaks of sentiment variation tend to appear just a minute or two before peaks of tweets.

Therefore, data suggest that monitoring sentiment during a match is a way to detect bursts of tweets just a couple minutes before they happen. Sudden sentiment variations even happen before any trend in the tweet volume time series is observable.

\subsection{Workload overview}

We used a set of games from the 2013 FIFA Confederations Cup to study the sentiment-volume relationship, derive statistics and models, and feed the sentiment analysis tool. The data is a set of dumps of tweets from 7 matches of the Brazilian soccer team: five matches from the FIFA Confederations Cup plus two friendly matches weeks prior the main event. The three first matches of the cup were for the group phase while the last two matches were the semi-final and the final. Table \ref{tab:tweet_volume} shows all matches and the total number of tweets read during the execution of the sentiment analysis tool.

\begin{table}
	\centering
	\caption{Matches information.}
	\begin{tabular}{ c c c c c}
		\hline
		\hline \textbf{BRA} & \textbf{Date} & \textbf{Total} & \textbf{Length} & \textbf{Tweets}		\vspace{-0.8mm}\\

		\textbf{vs} & & \textbf{tweets} & \textbf{(hours)} & \textbf{per hour}\\
		\hline 
		England & June 2nd & 370,471 & 2.62 & 141,401\\
		France & June 9th & 281,882 & 2.93 & 96,205\\
		Japan & June 15th & 736,171 & 4.08 & 180,434\\
		Mexico & June 19th & 615,831 & 3.79 & 162,488\\
		Italy & June 22nd & 518,952 & 3.42 & 151,740\\
		Uruguay & June 26th & 1,763,353 & 3.44 & 512,602\\
		Spain & June 30th & 4,309,863 & 4.18 & 1,031,067\\
		\hline 
		\hline
	\end{tabular}
	\label{tab:tweet_volume}
\end{table}

\begin{figure*}
	\centering
	\begin{subfigure}{0.24\textwidth}
		\centering
		\includegraphics[width=\linewidth]{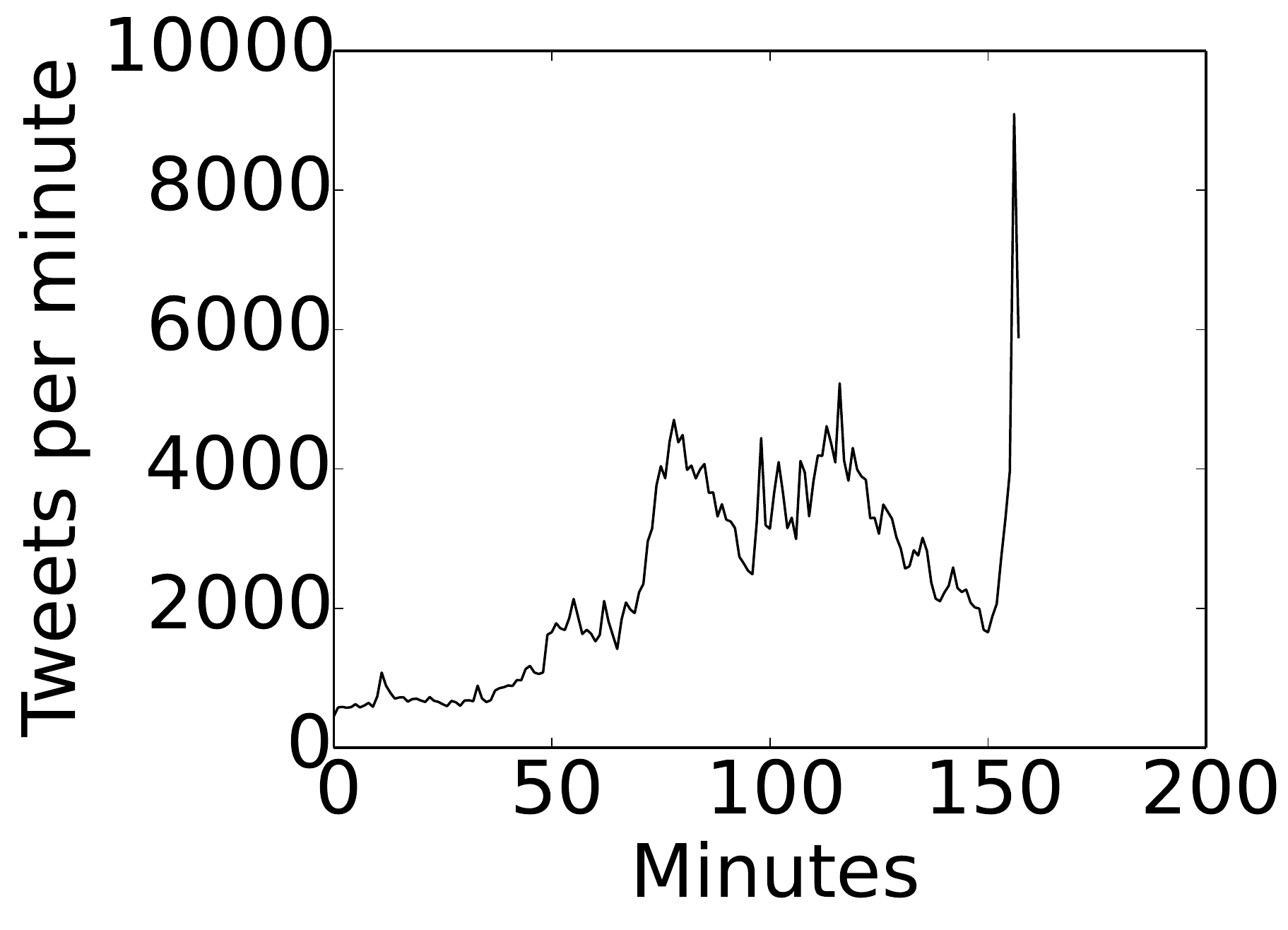}
		\caption{Brazil vs England}
		\label{fig:tweet_volume_BRAvING}
	\end{subfigure}
	\begin{subfigure}{0.24\textwidth}
		\centering
		\includegraphics[width=\linewidth]{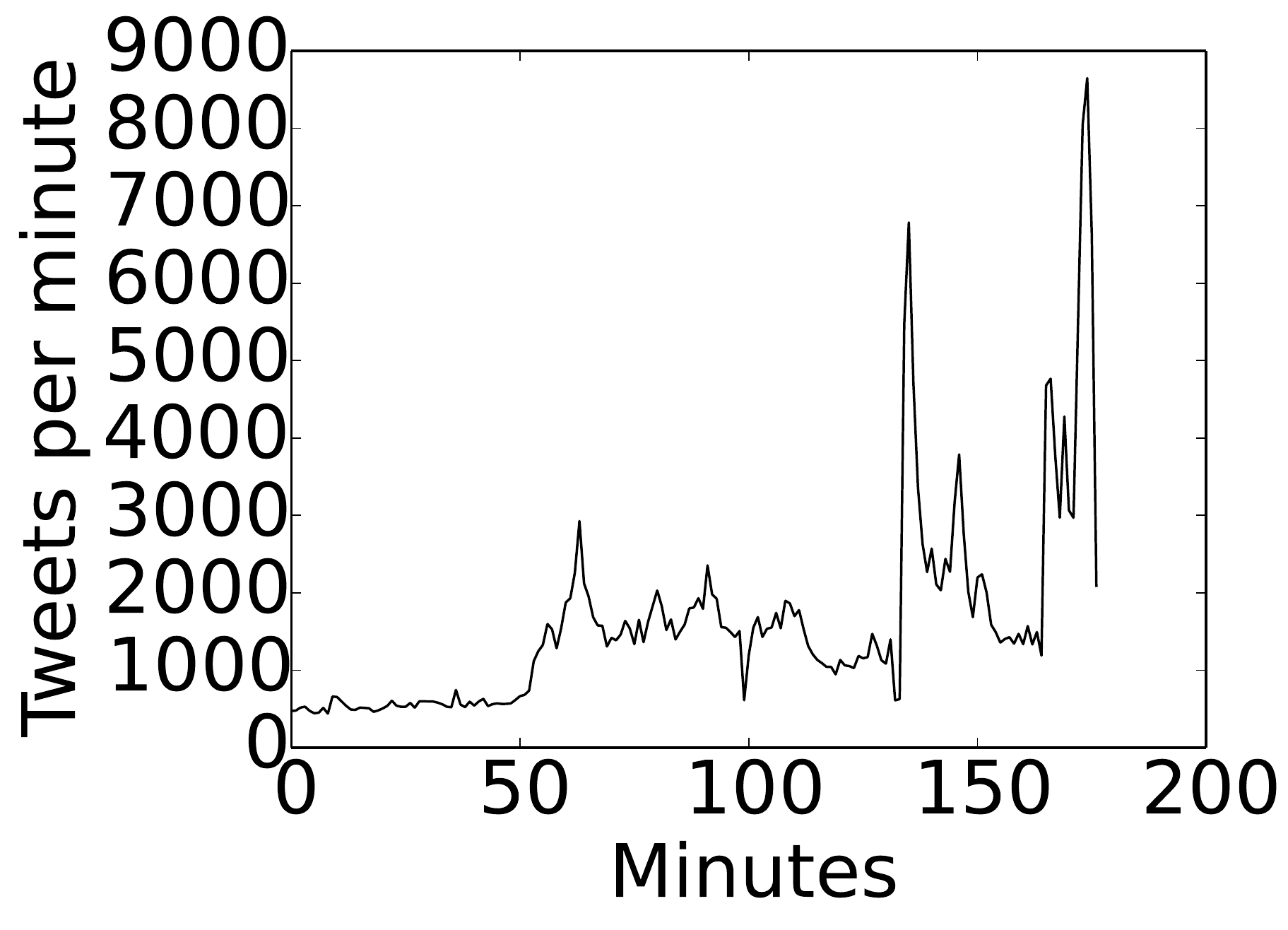}
		\caption{Brazil vs France}
		\label{fig:tweet_volume_BRAvFRA}
	\end{subfigure}
	\begin{subfigure}{0.24\textwidth}
		\centering
		\includegraphics[width=\linewidth]{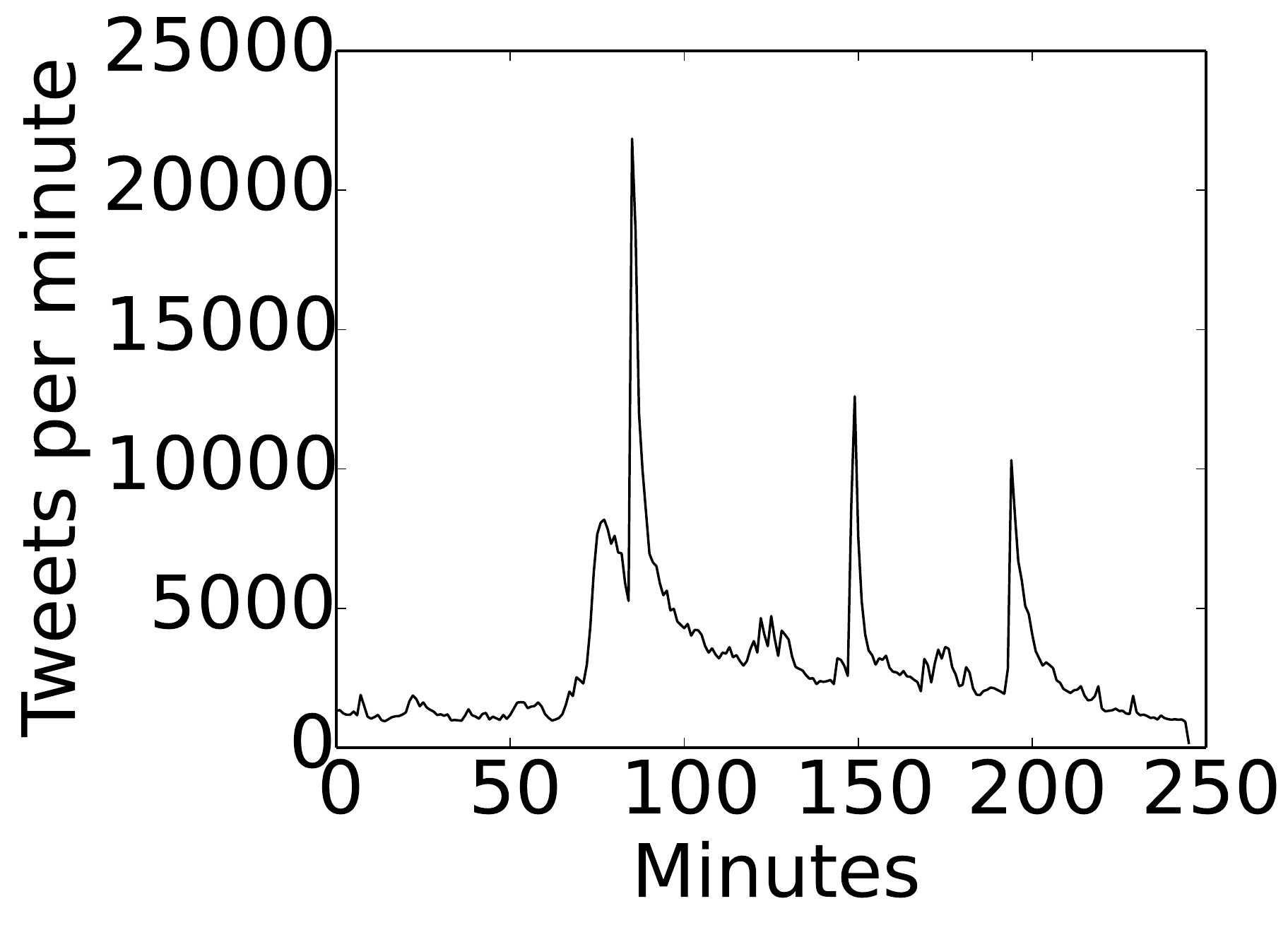}
		\caption{Brazil vs Japan}
		\label{fig:tweet_volume_BRAvJPA}
	\end{subfigure}
	\begin{subfigure}{0.24\textwidth}
		\centering
		\includegraphics[width=\linewidth]{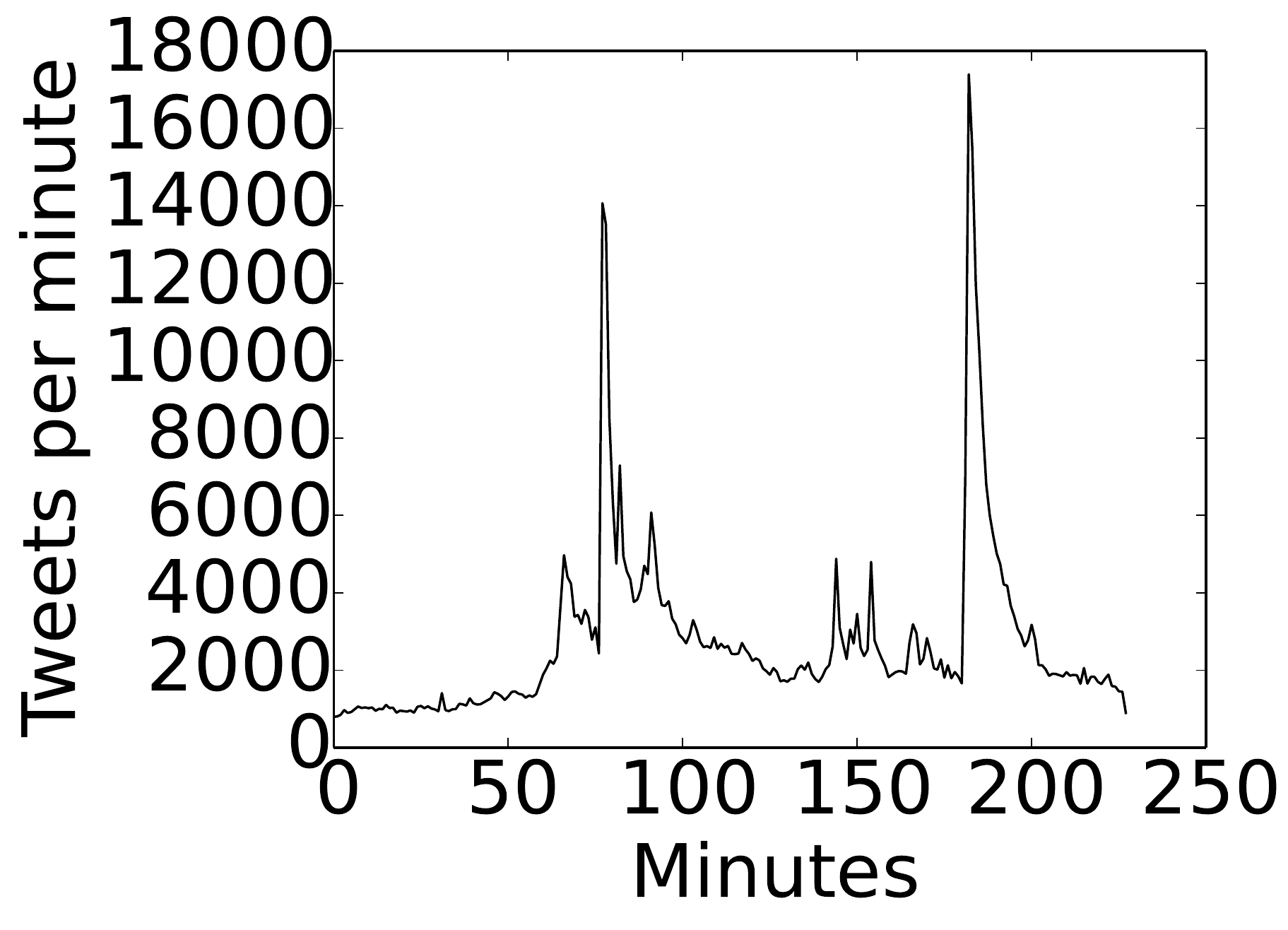}
		\caption{Brazil vs Mexico}
		\label{fig:tweet_volume_BRAvMEX}
	\end{subfigure}
	\\
	\begin{subfigure}{0.24\textwidth}
		\centering
		\includegraphics[width=\linewidth]{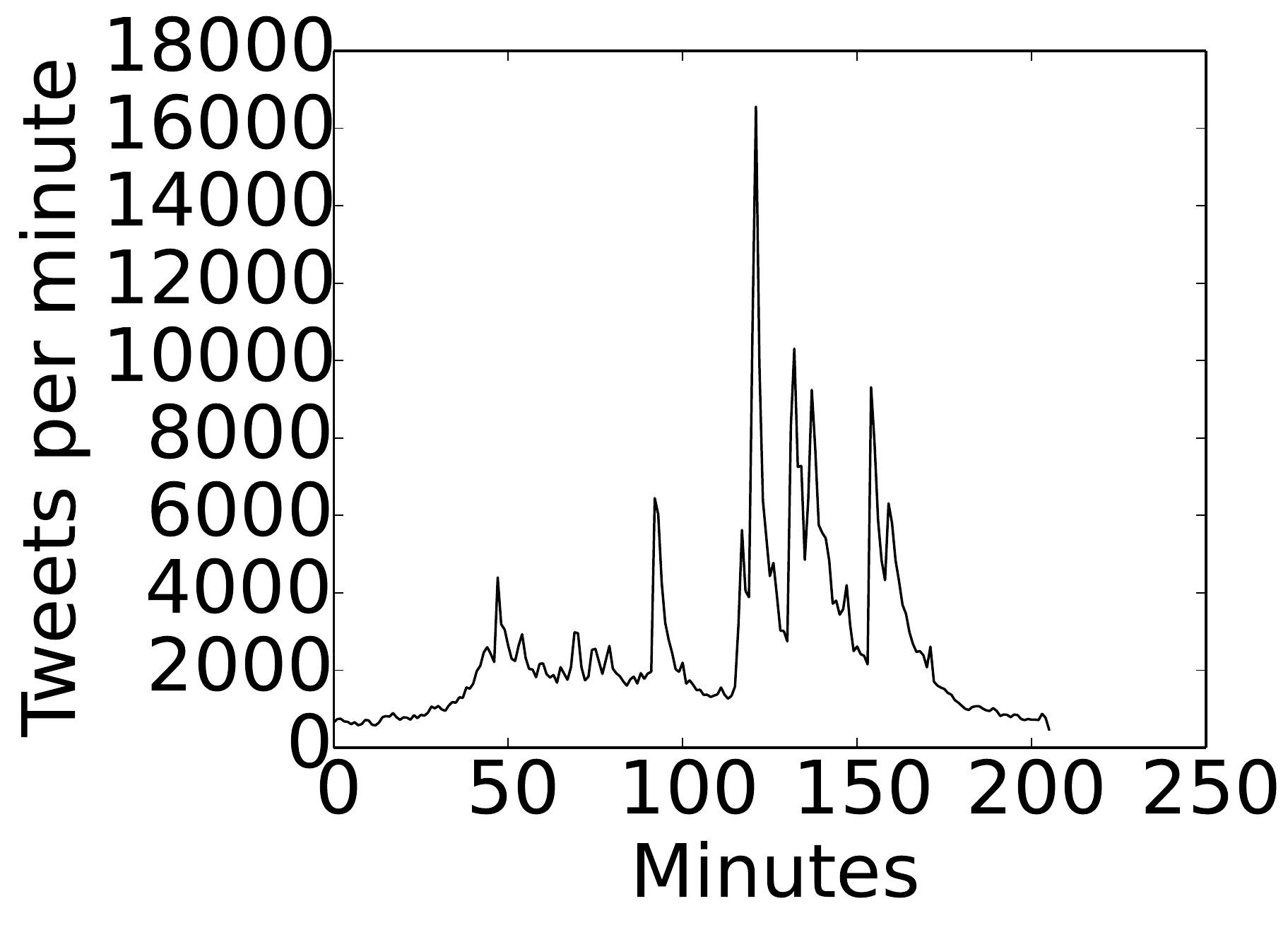}
		\caption{Brazil vs Italy}
		\label{fig:tweet_volume_BRAvITA}
	\end{subfigure}
	\begin{subfigure}{0.24\textwidth}
		\centering
		\includegraphics[width=\linewidth]{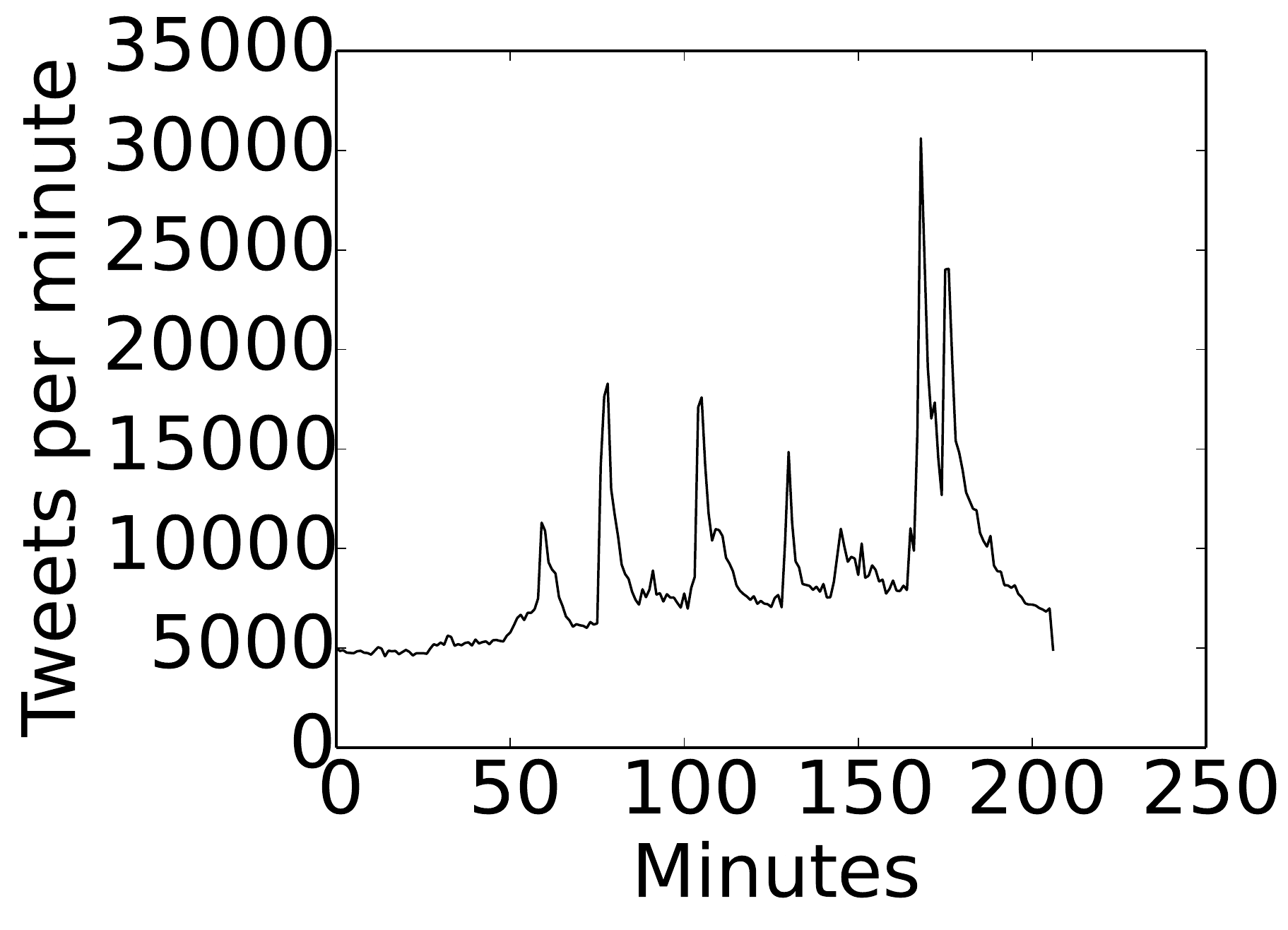}
		\caption{Brazil vs Uruguay}
		\label{fig:tweet_volume_BRAvURU}
	\end{subfigure}
	\begin{subfigure}{0.24\textwidth}
		\centering
		\includegraphics[width=\linewidth]{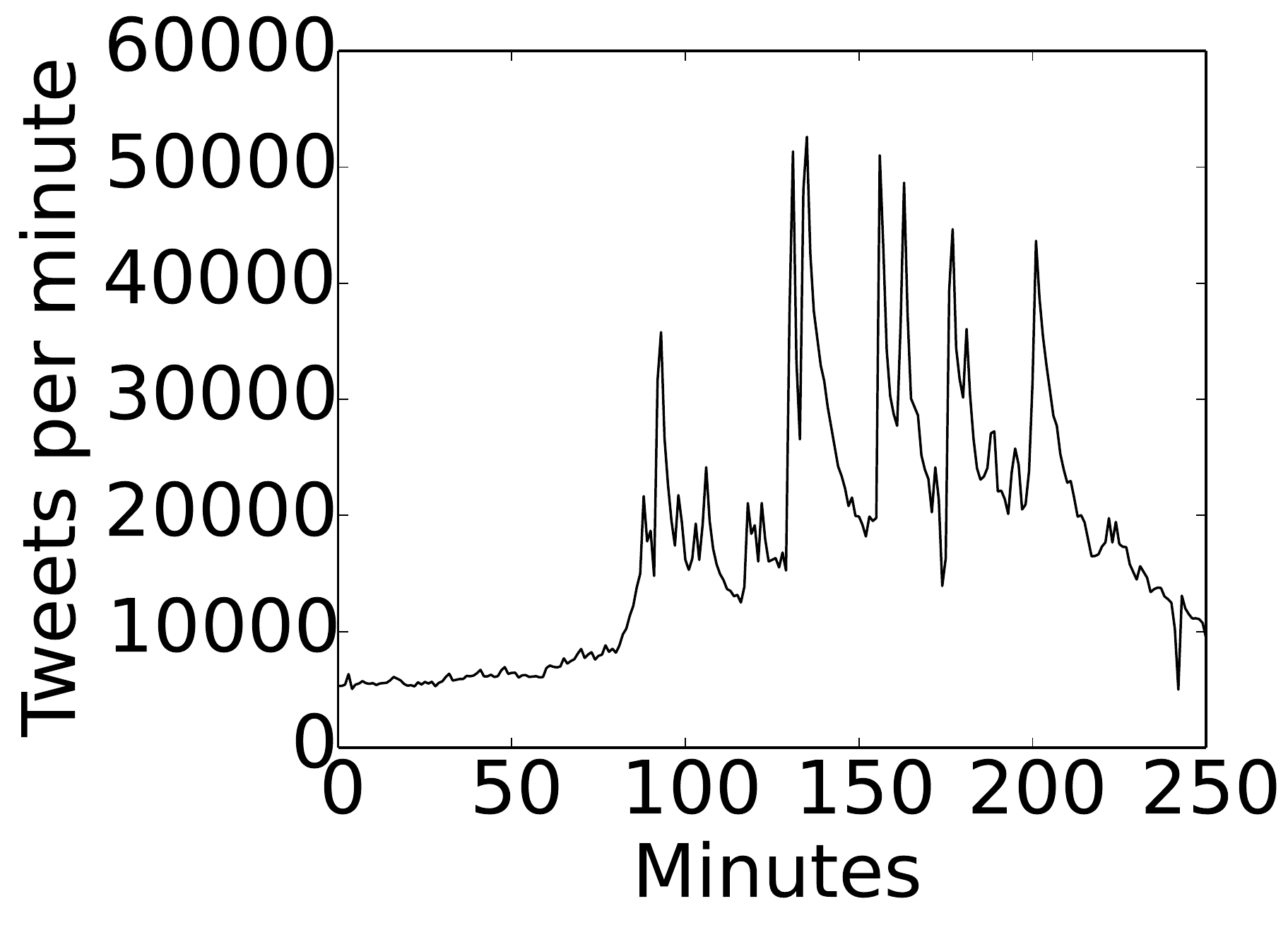}
		\caption{Brazil vs Spain}
		\label{fig:tweet_volume_BRAvESP}
	\end{subfigure}
	\caption{Tweets captured during the seven matches.}
	\label{fig:tweet_volume}
\end{figure*}

The two friendly matches were the ones with fewer tweets. They were also monitored for shorter periods of time. When the Confederations Cup began, games were monitored for longer and users initially showed more interest in tweeting about the games. Figure \ref{fig:tweet_volume} shows the time series for the volumes of tweets captured for all matches.

Time series peaks indicate a sudden increase in user interest on the match and are normally a consequence of notorious events. Experience has shown that polemical events like a goal saved on the last second generate more tweets than goals.

Both friendly games have peaks only close to the end of the monitoring, indicating that those games did not have much repercussion among social network users. Later games show more peaks during the match, reflecting the user enthusiasm increasing as the cup advances.

%% file: simulator.tex
\section{Evaluation Tool and Auto-scaling Triggers}
\label{sec:simulator}

In order to evaluate several and repeatable scenarios with different computing configurations, we created a simulation tool based on the in-house application for sentiment analysis. This section describes how the tool was created and validated and also the auto-scaling algorithms that were based on system and application metrics.

\subsection{Simulator}

A stream computing application can intuitively be thought as a network of queues, like the classic M/M/1, with a queue for each operator. But modeling each node in this network would require a great amount of effort and would possibly lead to very different behaviors than those on the original matches.

The main purpose of the simulator is to test new auto-scaling techniques on real world scenarios. Therefore only a limited randomness is desired to differentiate simulation from real matches. So instead of building a full featured sentiment-analysis application over Streams simulator, the idea is to randomize only the processing delay of the tweets, not their volume or distribution.

A tracer was attached to the in-house sentiment analysis application's code to monitor how tweets move through the processors. It logged the tweet id and the clock every time a tweet was parsed and every time it was finished being processed by the sink. It also logged from which PE the tweet came before reaching the sink so it would be easy to know its class, \textit{i.e.} the path it took, and whether/where it was discarded.

To model the delay distributions of a real instance of the sentiment analysis application, a test-bed comprising of a PC with 2.6 GHz CPU and 1 GB memory was used. The application was slightly adapted to read tweets from the dumps instead of Twitter. This way, the system could read all tweets at once and process them as fast as its CPU was able to. The memory was enough for the application and no other storage was used during runtime.

\begin{figure}
	\centering
	\includegraphics[width=0.9\linewidth]{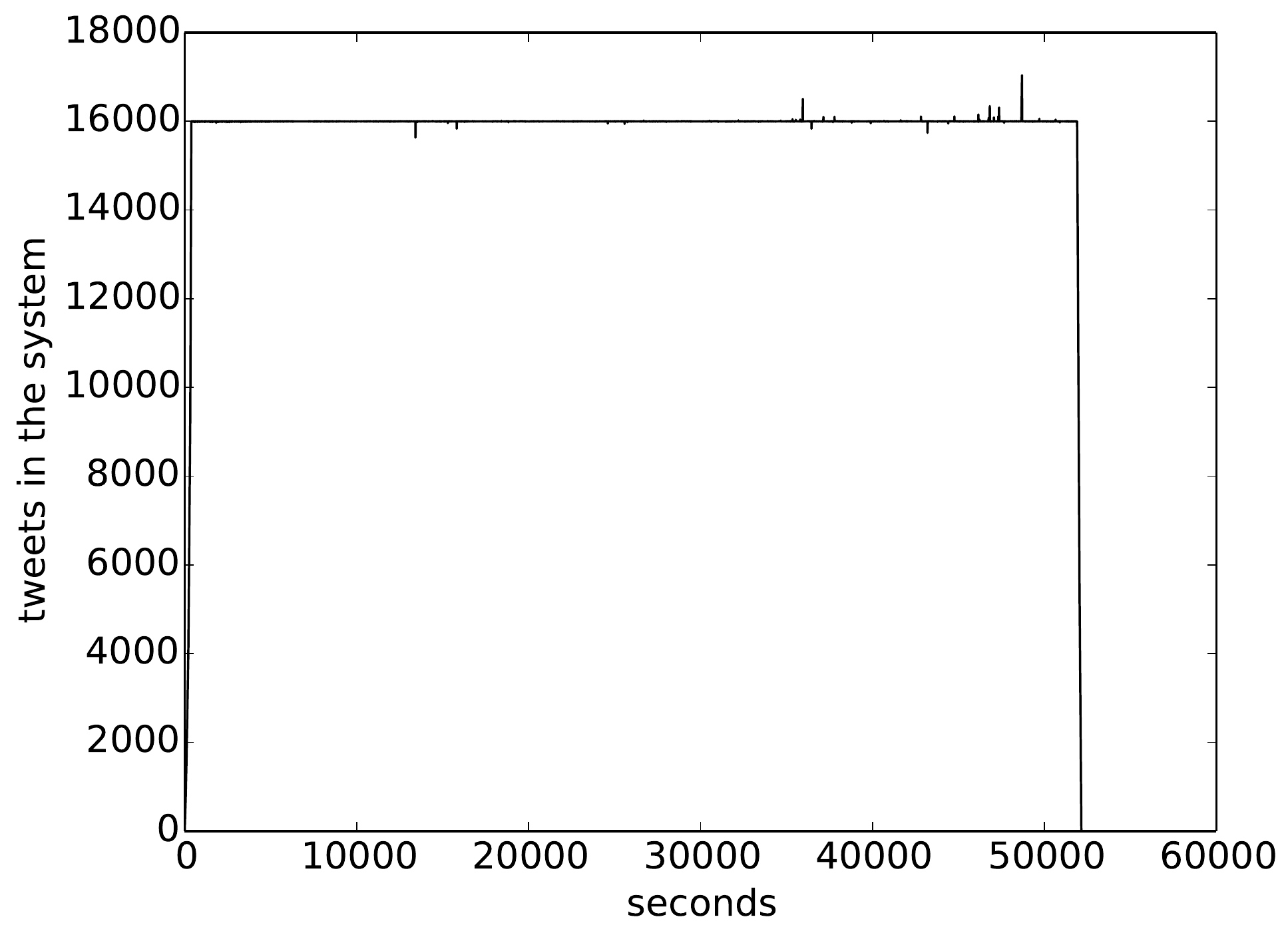}
	\caption{Number of tweets being processed simultaneously.}
	\label{fig:parallel_tweets}
\end{figure}

One at a time, all seven dumps were given to the system and the same behavior was observed every time: an almost constant number of tweets was processed in the system simultaneously (Figure \ref{fig:parallel_tweets}). By sampling on 1-second windows, the average number of tweets processed by the system was $15,875.32$ with a standard deviation of $1,233.80$, the average processing delay was of $192.09$ seconds and the average input rate of $82.65$ tweets/second. These numbers closely follow Little's Law:

$ L = \lambda W $

where $ L = 15,875.32$

and $ \lambda \times W = 82.65 \times 192.09 = 15,876.24 $

\begin{figure}
	\centering
	\includegraphics[width=0.9\linewidth]{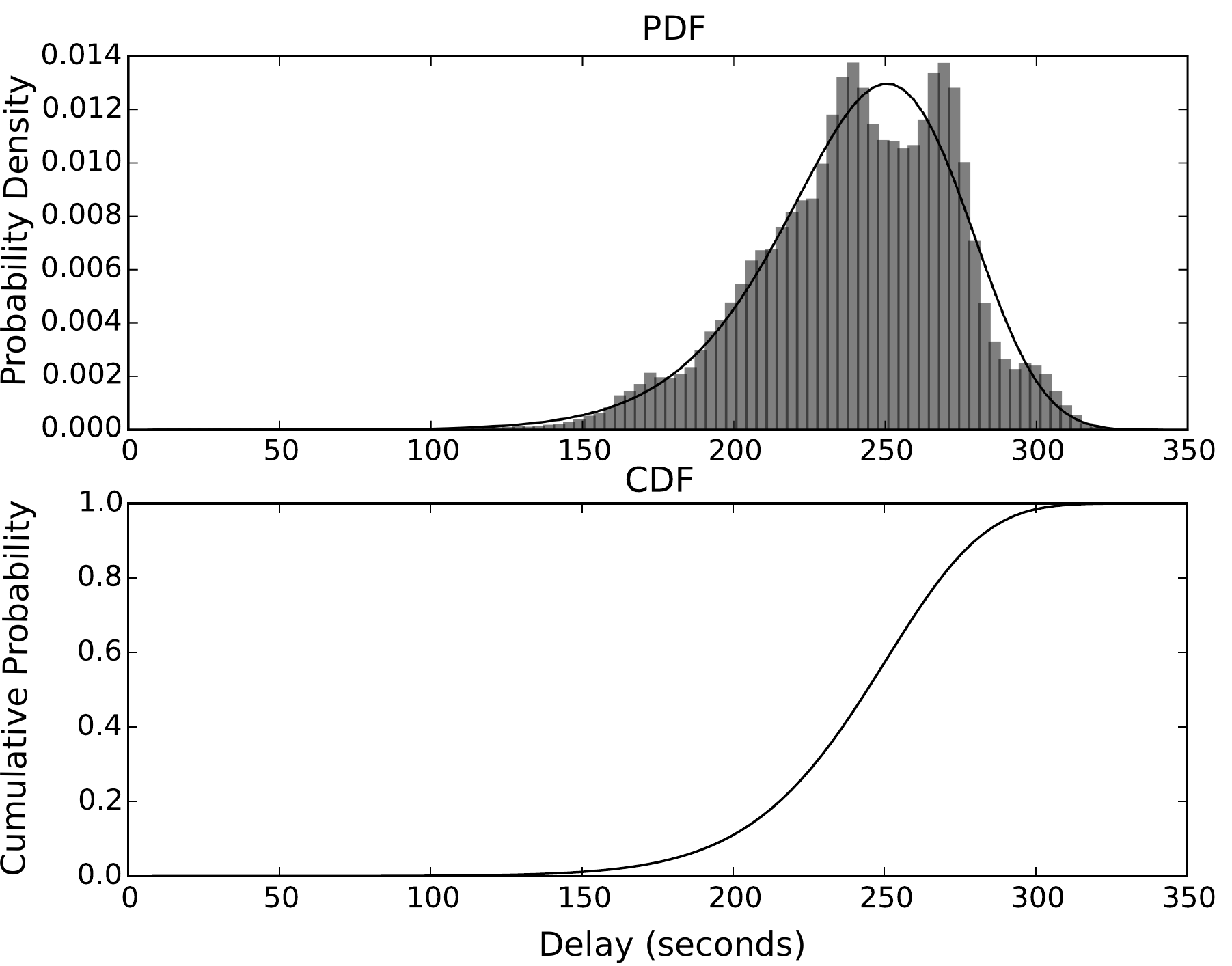}
	\caption{Weibull distribution of tweets with different topics.}
	\label{fig:histogram_TopicsFiltered}
\end{figure}

A trend is observable when grouping the tweets by their classes and analyzing their delay distribution. Figure \ref{fig:histogram_TopicsFiltered} shows the delay distribution of tweets that were considered off-topic and did not have their sentiment analyzed. After trying to fit different distributions to the histogram, the best match was the Weibull distribution with a normalized root mean square error of $0.01$.

Tweets that were discarded by PE (1) from Figure \ref{fig:fama_pipeline} had such a small delay (average below 1 second) that they were simply given a zero delay distribution in the simulator. As for the other paths, Weibull was also the best fit.

CPU utilization by the Streams process averaged 97.95\% and its memory usage averaged $812,044$ MB. From those statistics and the fact that Streams showed a predictable behavior while processing tweets in parallel, if it is assumed that CPU cycles are uniformly distributed to the tweets, there is a reasonable way to convert those delay distributions to CPU cycles distributions. That allows the extrapolation of the experiments to other machine configurations, making it possible to simulate any number of CPU frequencies and cores.

\subsection{Simulator internals}

For the simulations, tweet data from different sources was consolidated into a CSV file for each match. From the dump JSON files came the tweet id and post time. From the real processing in the sentiment analysis application came the tweet's class, processing delay and the sentiment score. Only the class and the processing delay are necessary for the derivation of the distribution parameters. The class, post time and sentiment scores were used for the simulations. Before the simulation begins all tweets are read from the CSV file and a random number of cycles is assigned to each tweet following its class distribution.

%The following scheme shows the fields: {\tt tweet id, post time, class id, processing delay, positive sentiment, negative sentiment, neutral sentiment.}

Unfortunately, running a discrete time simulator proved challenging on the algorithm complexity and a simpler discrete time model was adopted. This way, the simulator uses a certain time window on each iteration. By default, the simulation step is of one second. This means that all tweets that arrived during that time slot are read and CPU cycles available for a whole second are distributed among the current tweets.

The simulator has an internal clock that is incremented by the simulation step on each iteration of the main loop. The clock is initialized with the timestamp of the post time of the first tweet of the dump. Since it is not the objective to also simulate the network delay, a constant delay of zero is assumed and the tweet arrival time is considered equal to the post time.

% TODO: Recheck if we need this paragraph.
To simulate a limited input rate like Streams does, an input queue is used. All tweets posted during a simulation step are inserted on the queue, but only a configurable amount of tweets/second is read from the queue to be processed.

The beginning of the main loop is dedicated to reading all the tweets that were posted during that window. Tweets are read from the input queue respecting or not the input rate and are then stored in an internal processing structure where it will compete for resources.

Internally, this structure is a queue increasingly ordered by the post time. This helps the next part of the main simulation loop: distributing CPU cycles among the current tweets. If a tweet needs less cycles than there are available, excess cycles are equally distributed among the other current tweets. This is accomplished by using Algorithm \ref{alg:cpu_cyle}: 

\begin{algorithm}
	\caption{CPU cycle distribution algorithm.}
	\begin{algorithmic}
		\REQUIRE cyclesPerStep
		\REQUIRE tweetList

		\STATE numberOfCurrentTweets = length(tweetList)
		\STATE tweetsToProcess = numberOfCurrentTweets
		\STATE cyclesPerTweet = cyclesPerStep / numberOfCurrentTweets
		\STATE \textbf{sort} tweetList increasingly by remaining cycles
		\FOR {\textbf{each} tweet \textbf{in} tweetList}
			\IF {tweet.cyclesLeft $<$ cyclesPerTweet}
				\STATE excessCycles = cyclesPerTweet - tweet.cyclesLeft
				\STATE tweet.cyclesLeft = 0
				\STATE tweetsToProcess -= 1
				\STATE cyclesPerTweet += excessCycles / tweetsToProcess
			\ELSE
				\STATE tweet.cyclesLeft -= cyclesPerTweet
			\ENDIF
		\ENDFOR
	\end{algorithmic}
	\label{alg:cpu_cyle}
\end{algorithm}

The third part of the main simulation loop is getting rid of the tweets that are done being processed. Tweets that have used all cycles required are removed from that internal processing queue and are saved to a history log, from where statistics can later be taken: mean queue time, mean processing time, etc.

The last part is reacting to the current scenario by starting an up or downscale. This is not done on every simulation step, but rather only every few minutes. This adaptation frequency is configurable just as the provisioning time. For example, using the default values, every minute the situation is evaluated: sentiment and tweet volume for the last minutes are analyzed and a reaction might be issued for up or downscaling. After requesting or releasing resources, another amount of time will pass before they are available.

\subsection{Auto-scaling algorithms}
\label{sec:algorithms}

Two auto-scaling trigger algorithms are proposed based on \textit{a priori} knowledge of the application: 
\begin{enumerate}
\item \textbf{load algorithm:} knows the processing delay distributions of the sentiment analysis application;
\item \textbf{appdata algorithm}: only deals with peaks, is oblivious to ordinary increases of traffic and runs alongside the load algorithm---it uses the sentiment analysis data generated by the application itself.
\end{enumerate}

The \textit{load algorithm} is based on the expected time to process all current tweets versus the given SLA. The estimated delay is calculated from the quantile function of the delay distribution of the different tweet classes and from the proportion of the class length. The quantile value is a parameter to the simulator.

A quantile of $0.5$ is the median and roughly means a delay that is greater or equal to half of the observable delays. A quantile of $0.9$ will return a delay estimative that will cover as much as 90\% of the tweets. The higher the quantile the more pessimistic the model is and more likely it is to react before the SLA is really violated. On the other hand, a higher quantile will also spend more resources. Each class estimated delay is then weighted according to the class length known from the training data.

Since this algorithm is proposed as a simple reactive algorithm, no predictions on the future of tweet volume is attempted. Instead, if the expected delay is above the SLA, more resources are allocated, and if the expected delay is below half the SLA, resources are released. Downscaling is limited to a single CPU being returned at a time, so sudden increases in tweet volume have less impact. For upscaling, an estimate of necessary resources is calculated by the proportion of the expected delay and the SLA over the current available resources, as shown in the formula below:

$cpus_{nextPeriod} = ceil(cpus * (expectedDelay / SLA))$

The \textit{appdata algorithm} analyzes the average sentiment score of the last minutes and compares it to the average sentiment of the minutes before. If the sentiment score increases by $0.5$ or more, a predefined quantity of new CPUs is allocated.

The two proposed algorithms are used in opposition of the classic and largely adopted auto-scaling algorithm: the CPU usage \textbf{threshold algorithm}. The way this algorithm was implemented in the simulator, every time the average CPU usage goes above a certain predefined threshold, an extra CPU is allocated. On the other hand, every time the CPU usage is below 50\%, a CPU is released.

%% file: results.tex
\section{Simulation results and analysis}
\label{sec:results}

The goal of the experiments presented in this section is to compare the performance of the \textit{load} and \textit{appdata} algorithms against the classic CPU usage \textit{threshold} algorithm.

For the CPU usage \textit{threshold} algorithm, thresholds of 60\%, 70\%, 80\%, 90\%, and 99\% are used. For the \textit{load} algorithm, quantile values are: 90\%, 99\%, 99.9\%, 99.99\% and 99.999\%. The \textit{appdata} algorithm was run alongside the load algorithm with a quantile of 99.999\% and different values of extra CPUs allocated when peaks were detected: from 1 to 10.

All simulations were run with the configurations described in Table \ref{tab:scenario_config}. All scenarios were repeated until the length of the confidence interval with 95\% confidence was smaller than 10\% of the mean.

\begin{table}[!h]
	\centering
	\caption{Basic configuration for all simulation scenarios.}
	\begin{tabular}{ c c }
		\hline
		\hline
		\textbf{Variable} & \textbf{Value}\\
		\hline
		CPU frequency & 2.0 GHz \\
		starting CPUs & 1 \\
		simulation step & 1 second \\
		SLA & 300 seconds \\
		adapt frequency & 60 seconds \\
		resource allocation time & 60 seconds \\
		\hline
		\hline
	\end{tabular}
	\label{tab:scenario_config}
\end{table}

\subsection{Load algorithm performance}

\begin{figure*}
	\centering
	\begin{subfigure}{0.24\textwidth}
		\centering
		\includegraphics[width=\linewidth]{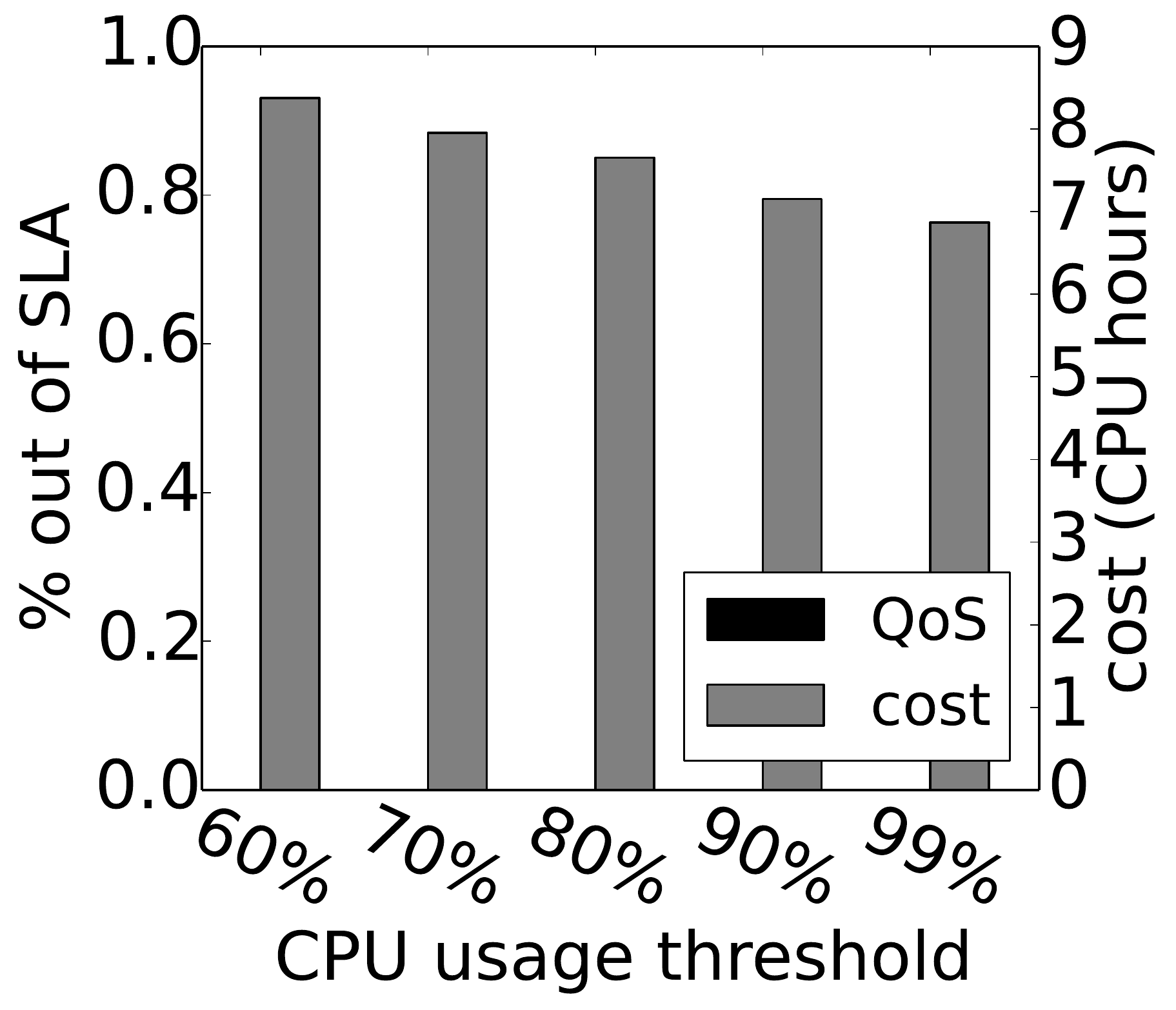}
		\caption{Threshold: Brazil vs Japan}
		\label{fig:bar_BRAvJPA_threshold}
	\end{subfigure}
	\begin{subfigure}{0.24\textwidth}
		\centering
		\includegraphics[width=\linewidth]{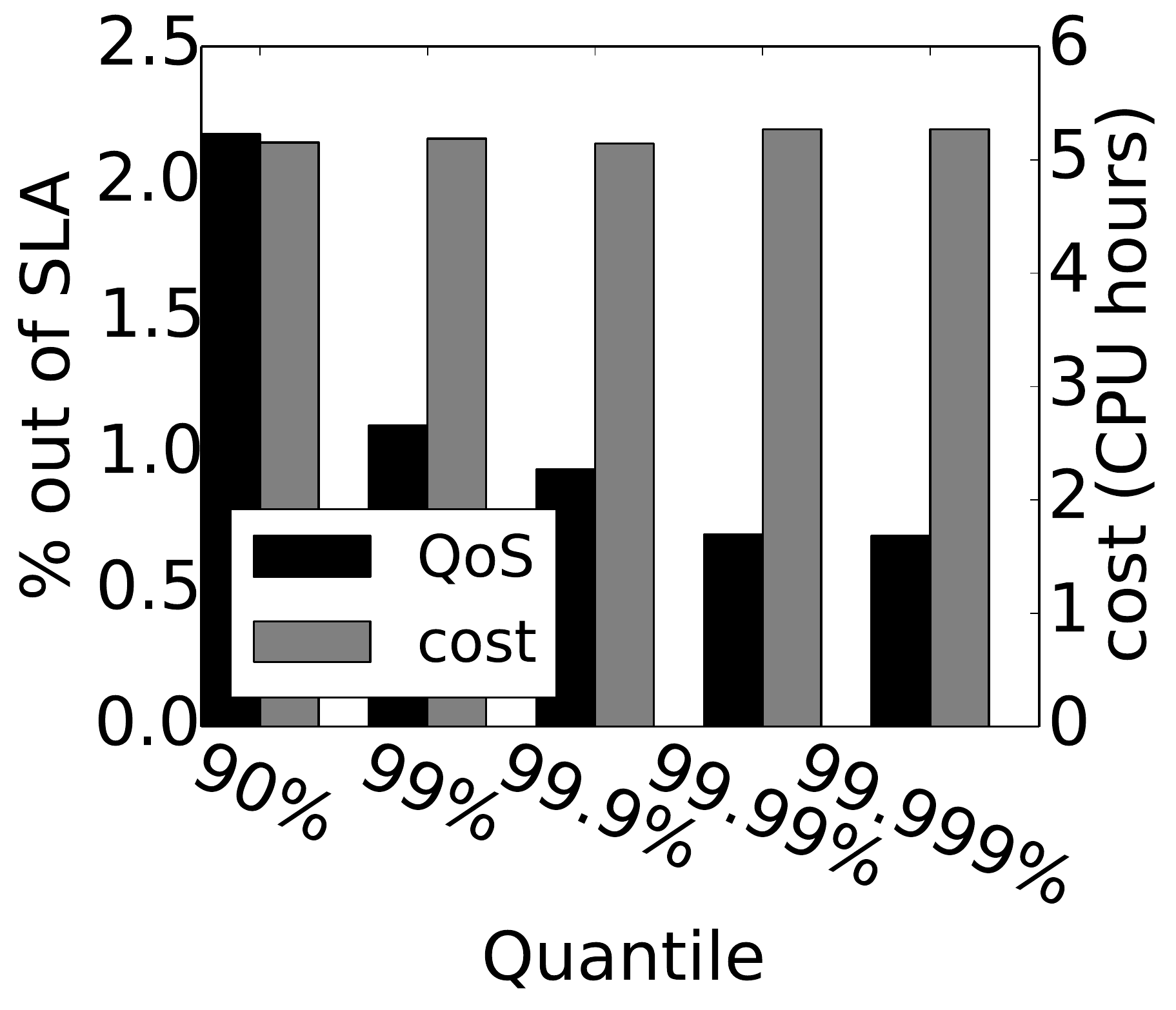}
		\caption{Load: Brazil vs Japan}
		\label{fig:bar_BRAvJPA_load}
	\end{subfigure}
	\begin{subfigure}{0.24\textwidth}
		\centering
		\includegraphics[width=\linewidth]{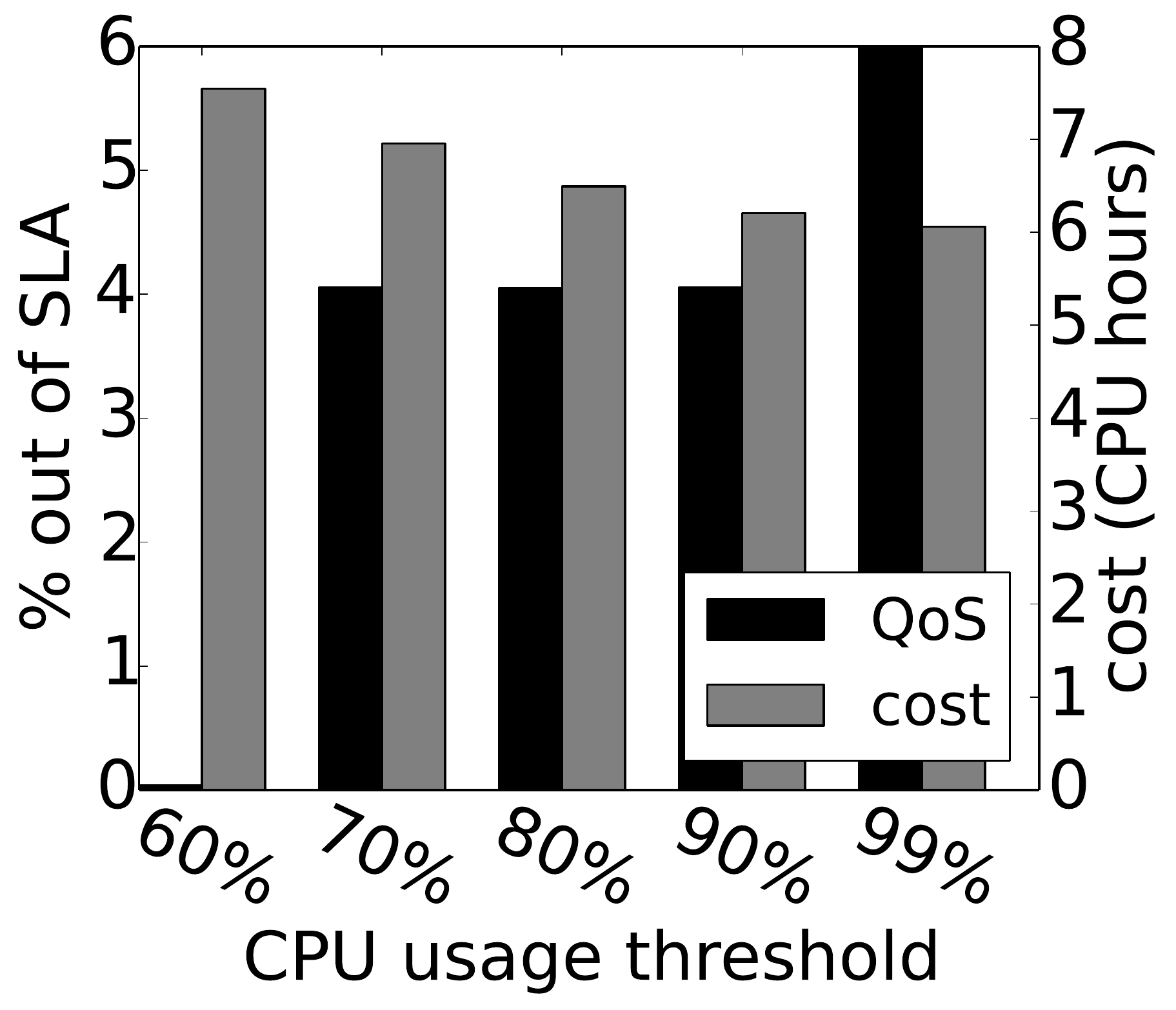}
		\caption{Threshold: Brazil vs Mexico}
		\label{fig:bar_BRAvMEX_threshold}
	\end{subfigure}
	\begin{subfigure}{0.24\textwidth}
		\centering
		\includegraphics[width=\linewidth]{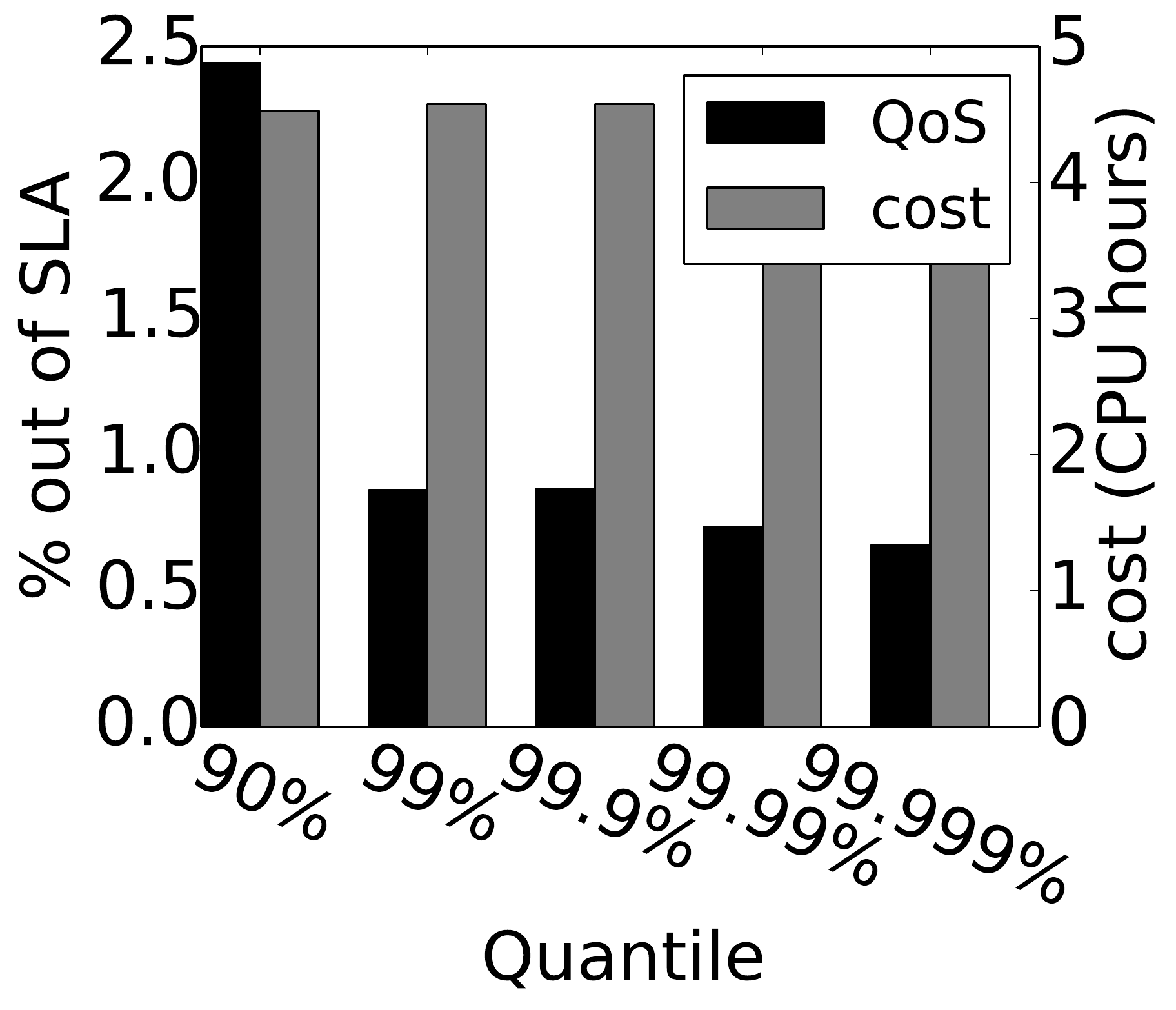}
		\caption{Load: Brazil vs Mexico}
		\label{fig:bar_BRAvMEX_load}
	\end{subfigure}
	\\
	\begin{subfigure}{0.24\textwidth}
		\centering
		\includegraphics[width=\linewidth]{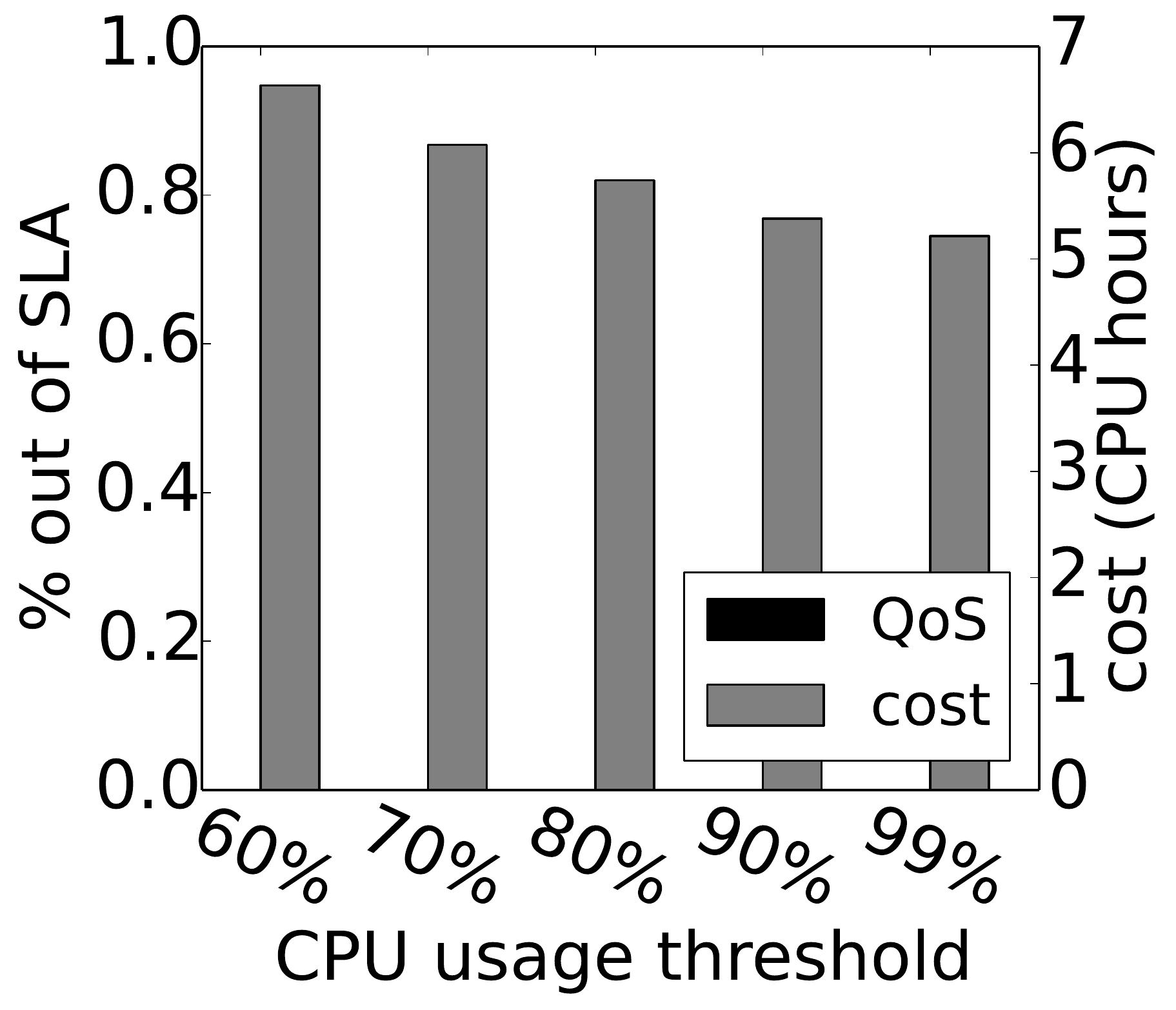}
		\caption{Threshold: Brazil vs Italy}
		\label{fig:bar_BRAvITA_threshold}
	\end{subfigure}
	\begin{subfigure}{0.24\textwidth}
		\centering
		\includegraphics[width=\linewidth]{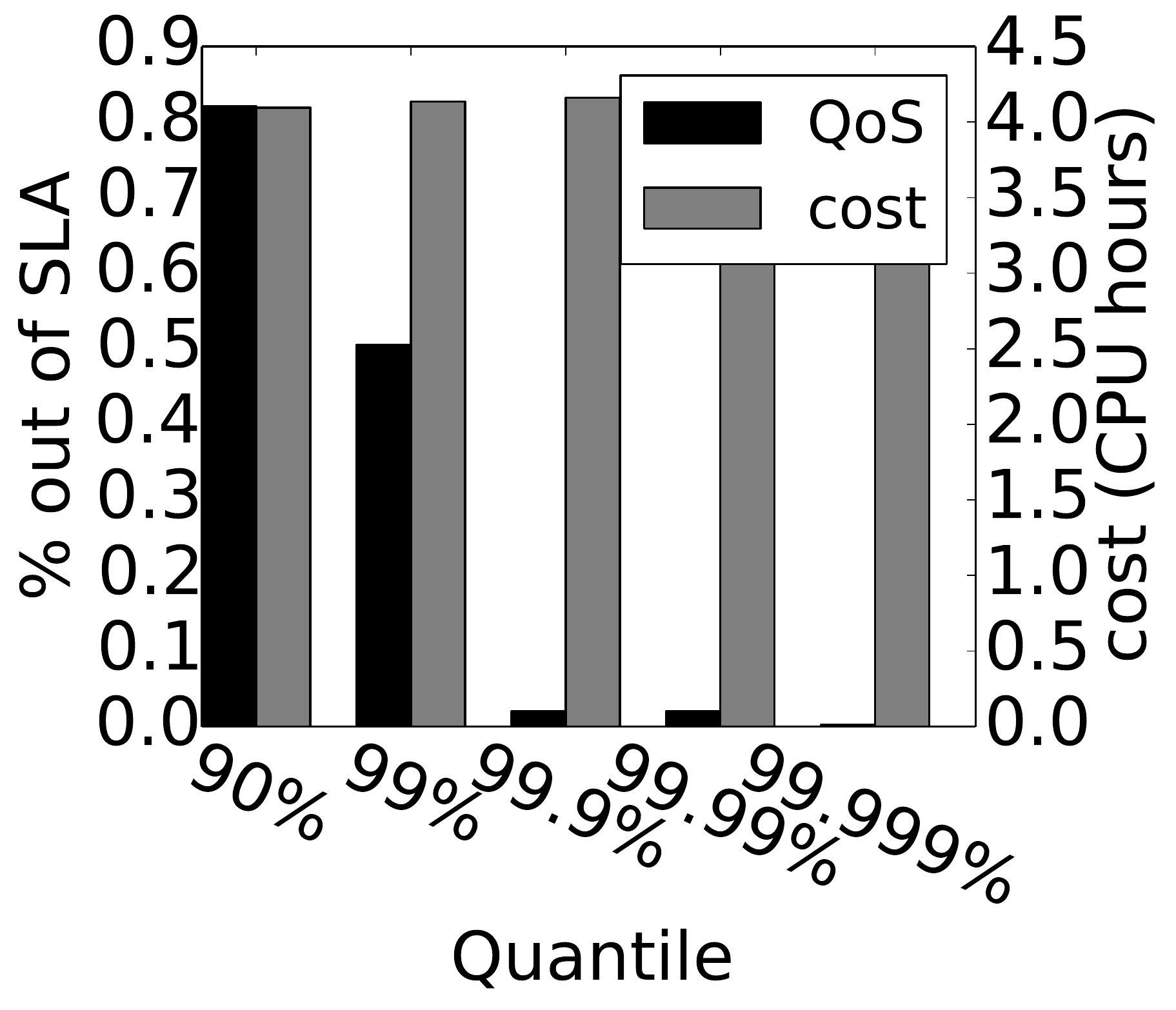}
		\caption{Load: Brazil vs Italy}
		\label{fig:bar_BRAvITA_load}
	\end{subfigure}
	\begin{subfigure}{0.24\textwidth}
		\centering
		\includegraphics[width=\linewidth]{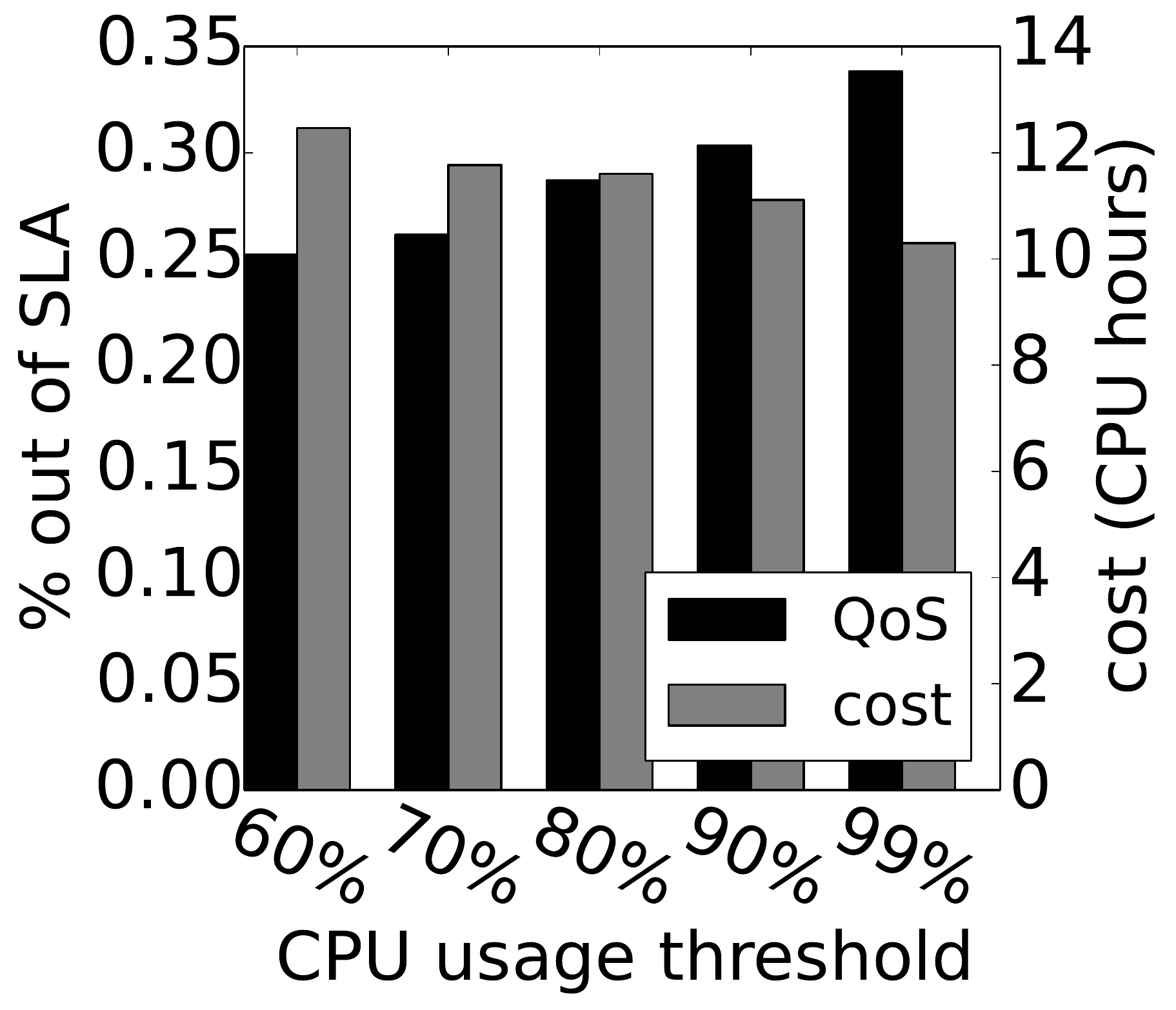}
		\caption{Threshold: Brazil vs Uruguay}
		\label{fig:bar_BRAvURU_threshold}
	\end{subfigure}
	\begin{subfigure}{0.24\textwidth}
		\centering
		\includegraphics[width=\linewidth]{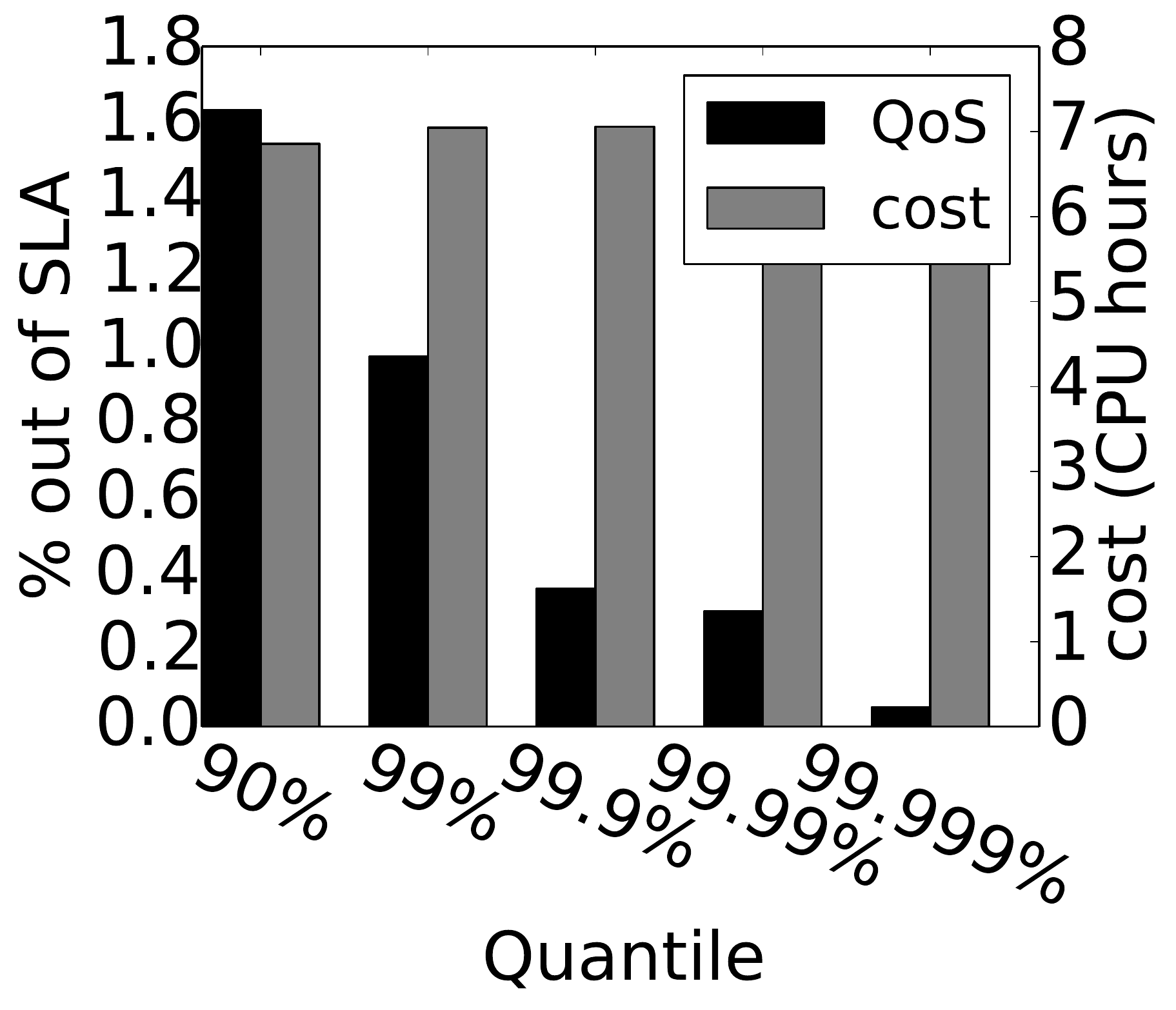}
		\caption{Load: Brazil vs Uruguay}
		\label{fig:bar_BRAvURU_load}
	\end{subfigure}
	\\
	\begin{subfigure}{0.24\textwidth}
		\centering
		\includegraphics[width=\linewidth]{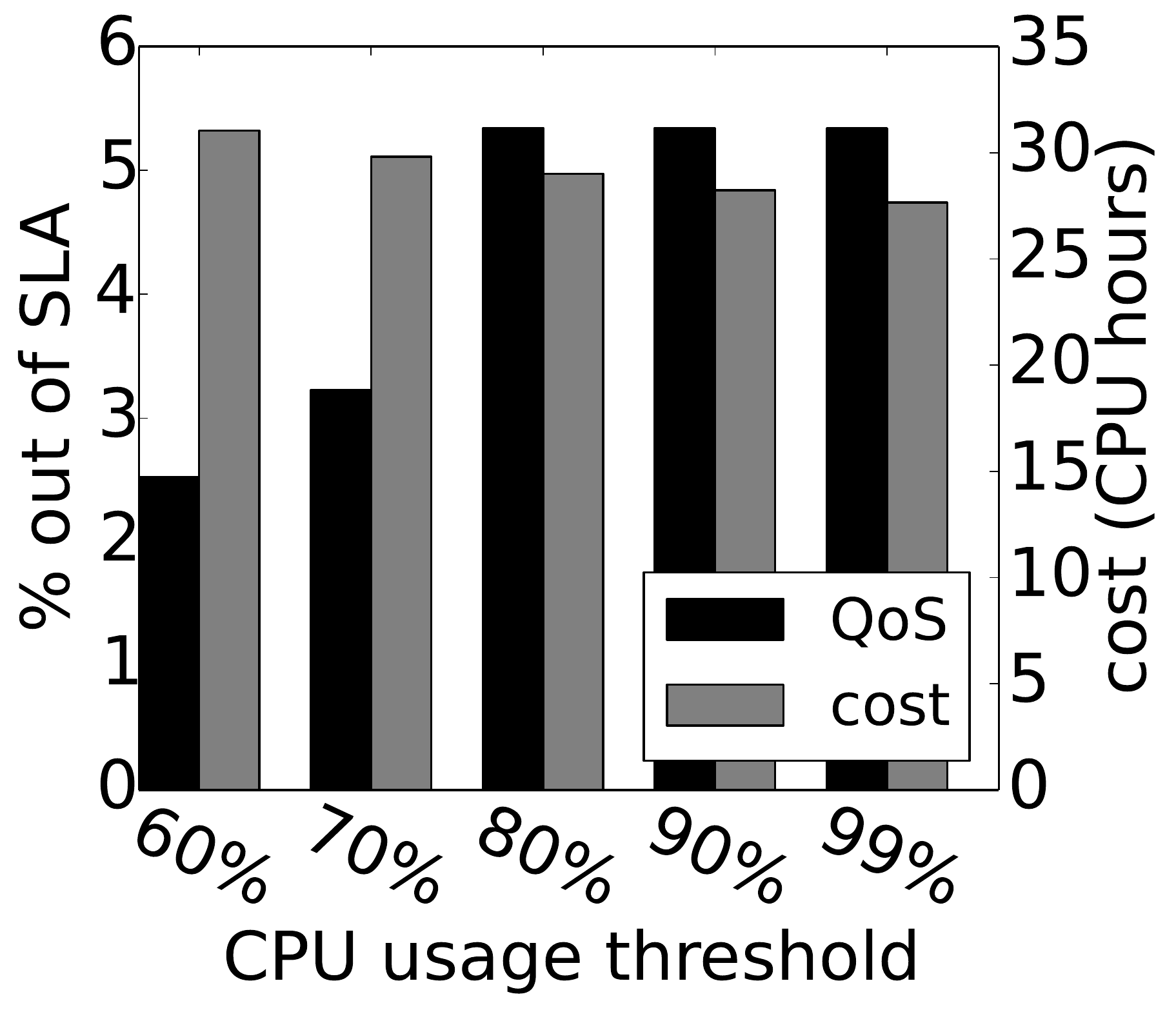}
		\caption{Threshold: Brazil vs Spain}
		\label{fig:bar_BRAvESP_threshold}
	\end{subfigure}
	\begin{subfigure}{0.24\textwidth}
		\centering
		\includegraphics[width=\linewidth]{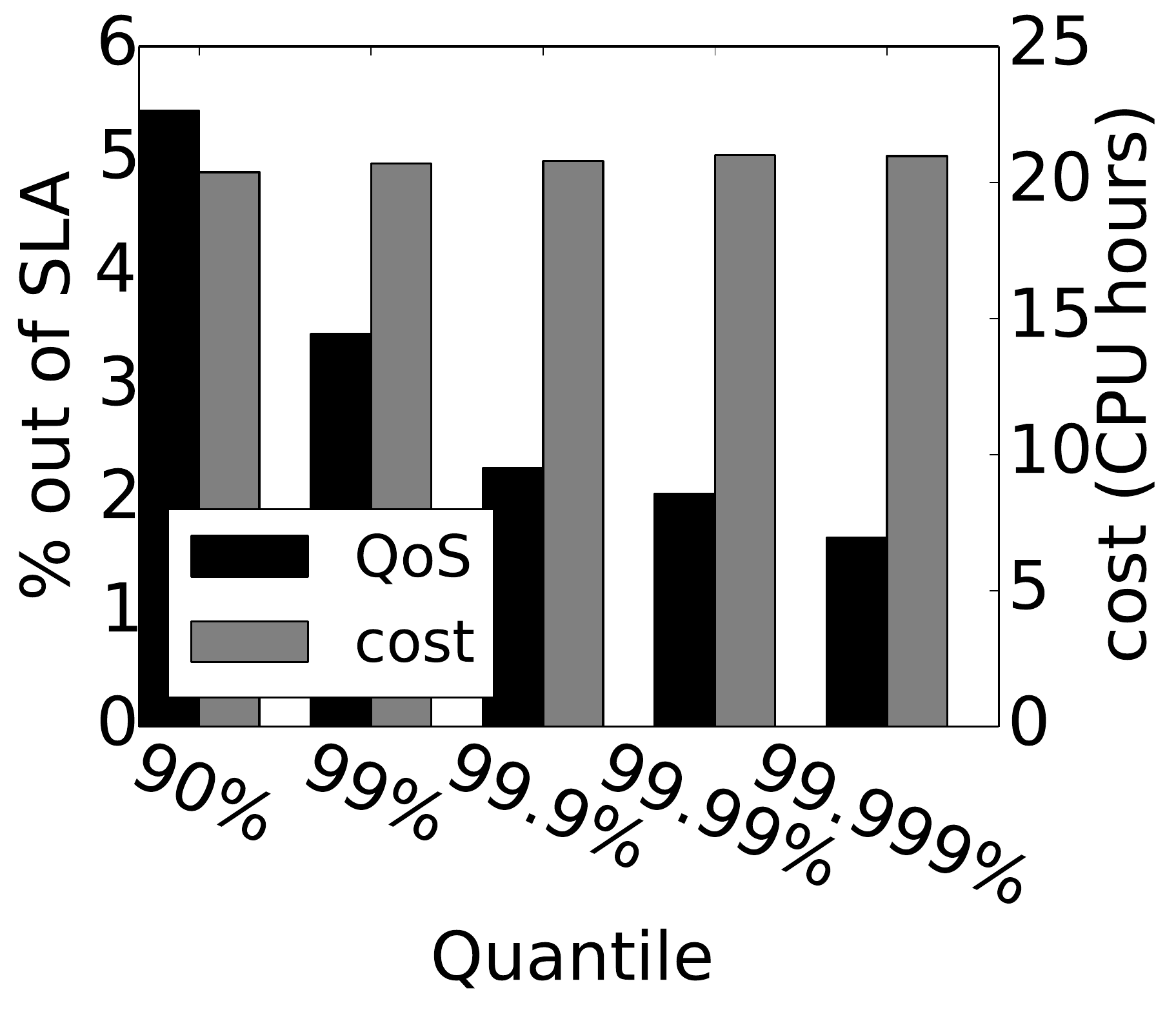}
		\caption{Load: Brazil vs Spain}
		\label{fig:bar_BRAvESP_load}
	\end{subfigure}
	\caption{Comparison of the performance of \textit{threshold} and \textit{load} algorithms for five of the seven games.}		\label{fig:matches_quality_cost}
\end{figure*}

Simulations were first run to compare the performance in terms of quality and cost of the load algorithm and the classic CPU usage threshold algorithm. Figure \ref{fig:matches_quality_cost} was built from the resulting data and shows the evolution of the quality and the cost of each match as a function of the algorithms and parameters. Quality is shown in terms of percentage of tweets that took longer than the SLA requirement to be processed.

Matches of Brazil against England and France were left out of the figure as there was close to no difference on the algorithms to be shown. On those matches, the volume of tweets was not as significant as on the other matches which made it easier for the auto-scaling algorithms to react to the relatively small variations of volume. In fact, both the threshold and the load algorithms performed perfectly for both matches and not a single tweet took longer than the SLA to be processed on all simulated scenarios.

The load algorithm had a fairly constant cost among all used quantiles: $2.76$ CPU hours were used for the England match and $3.03$ CPU hours for the France match. Cost differences for different quantiles is insignificant. This behavior repeats on all seven matches as seen in Figure \ref{fig:matches_quality_cost} and shows how predictable the algorithm is in terms of cost.

The threshold algorithm is more expensive for both matches ranging from $3.48$ CPU hours (threshold of 99\%) to $4.52$ CPU hours (threshold of 60\%) for the match with England and $3.41$ (threshold of 99\%) to $3.96$ (threshold of 60\%) for the match with France. The cost as a function of the CPU usage threshold always decreases as the threshold increases, as is observable on all other matches shown on Figure \ref{fig:matches_quality_cost}.

The three matches of the group phase showed close patterns and volumes of tweets. But while the threshold algorithm was able to perform perfectly for the matches of Japan and Italy, it did not show the same performance for the Mexico match. For this match, only a threshold of 60\% CPU usage was close to completely meeting the SLA with only 0.04\% of tweets above the target processing time.

For those three matches, the load algorithm was able to perform well although not perfectly on the quality side. In general terms, the higher the quantile used, the best the algorithm performs with an insignificant increase in cost. The load algorithm was able to always deliver lower costs, an advantage that is present on every simulated scenario. Nevertheless, the load algorithm was able to perform better than the threshold algorithm for the Mexico match.

The reason for the generally better performance of the load algorithm on the Mexico game is the great peak of tweets that happens around 180 minutes of the monitoring of the match (refer to Figure \ref{fig:tweet_volume}). Even if the peak does not seem very different from other peaks of the other matches, it happens more abruptly while others have small increase just before.

The load algorithm performs better because it has the ability to upscale the number of CPUs faster. While the threshold algorithm can only increase the number of CPUs by one per observation, the other algorithm increases by as many times as the proportion of the estimated delay and the SLA (as seen on Section \ref{sec:algorithms}), an ability that comes from the \textit{a priori} knowledge of the delay distribution. Those peaks are events the threshold and the load algorithms were not designed to deal with and the reason the appdata algorithm is proposed.

The last two matches had by far more tweets and also more significant peaks. None of the two algorithms performed perfectly for them, but this time the load algorithm performed significantly better when configured with higher quantiles while using way less resources. Those two matches were specially challenging for the algorithms thanks to the large amounts and great bursts of tweets posted by the fans that were watching the final games of the championship. While the threshold algorithm was still able to perform reasonably for the Uruguay match, the final match had the highest number of peaks of all games and the load algorithm capacity to upscale fast was decisive for making it outperform the threshold algorithm.

On the Brazil vs Uruguay match, comparing the scenario configurations with the best performances, the load algorithm with 99.999\% quantile delivered 0.05\% of tweets above the SLA while costing 7.14 CPU hours. The threshold algorithm with a 60\% CPU usage threshold had 0.25\% of the tweets missing the SLA at a cost of 12.46 CPU hours. For the final match against Spain and the same scenarios, the load algorithm had 1.67\% of tweets above the SLA with a cost of 20.97 CPU hours while the threshold algorithm let 2.52\% of the tweets lose the SLA with a cost of 31.04 CPU hours.

For the Brazil vs Uruguay match, replacing the traditional threshold algorithm with a 60\% threshold by the load algorithm means saving 43\% CPU hours with a slight improve of quality. For the Spain game, savings are of 33\%. It is important, however, to note that rarely such a low threshold is used on ordinary jobs on the cloud.

\subsection{Appdata algorithm performance}

The appdata algorithm detects peaks through the analysis of the live stream of sentiment taken from the tweets being processed. Its use was put to test together with the load algorithm with a 99.999\% quantile and a number of extra CPUs varying from 1 to 10.

As shown in Section \ref{sec:sentiment_analysis}, peaks of tweets can be detected by analyzing sudden changes in user sentiment. CPUs allocated preemptively are available when peaks occur and more resources are necessary, preventing quality loss. In that context, a window of 60 seconds is compared to a previous window of same size. Peaks are consequences of certain events and the first few tweets related to the event that come before the peak are the key to detecting them. Older tweets, from before the event, that just happened to take longer to process cannot be confused with those few first peak tweets even if they are done being processed at the same time. For this, care must be taken that it is not the time the tweet is done being processed that is used to analyze the sentiment time series, but the tweets post time.

In practice, windows of 60 seconds of length are not large enough for efficiently detecting peaks. If at a given time, only tweets that were posted at most 60 seconds sooner are considered for a window, very few will be taken into account as very few are done being processed under these 60 seconds. After testing different lengths of windows, the one that rendered the best results was the one of 120 seconds. With that size, even if most tweets are not done being processed, a sufficiently large number of tweets with sentiment are available for detecting peaks.

\begin{figure}
	\centering
	\includegraphics[width=\linewidth]{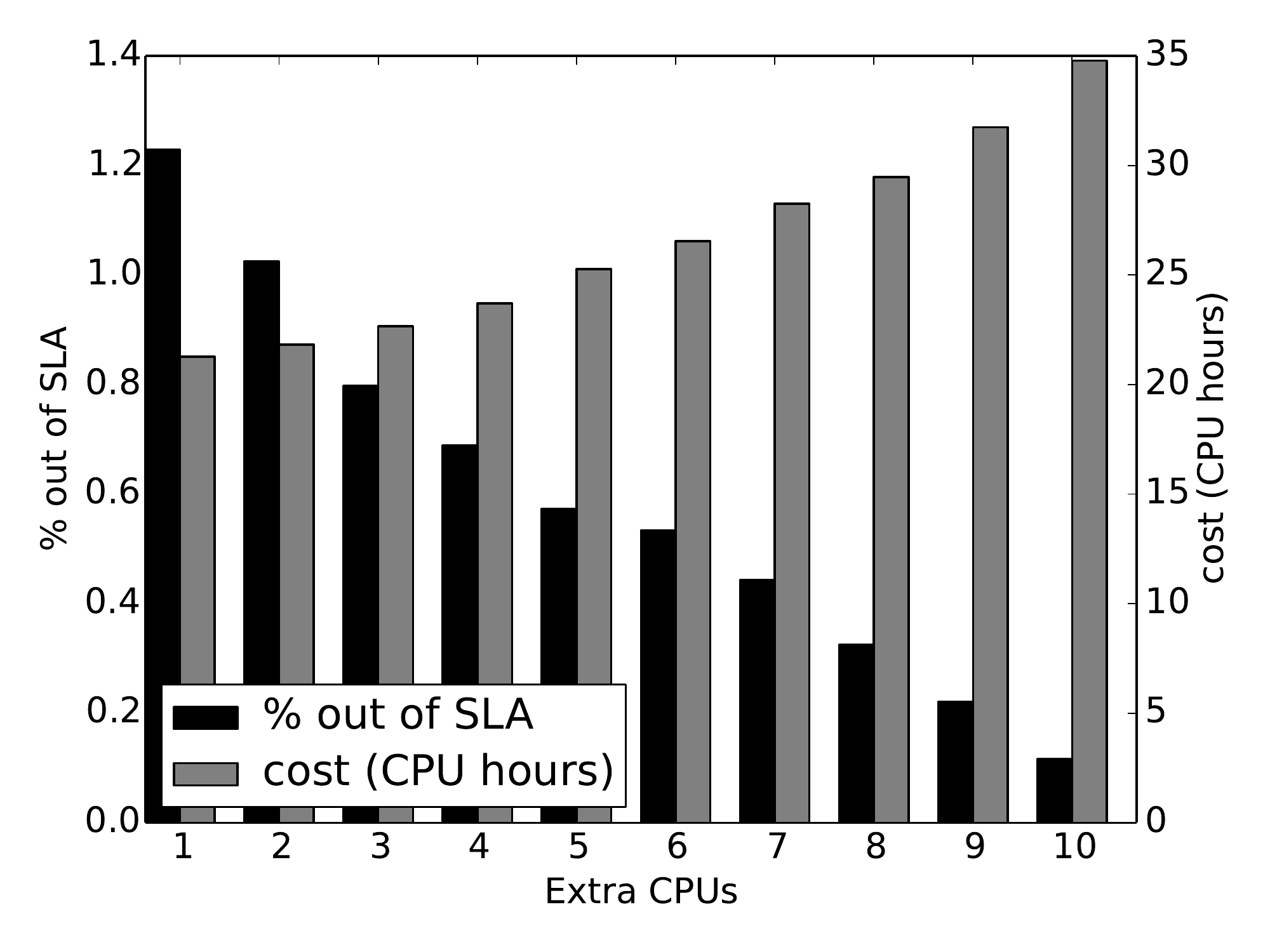}
	\caption{Appdata: Brazil vs Spain.}
	\label{fig:bar_BRAvESP_peak}
\end{figure}

Figure \ref{fig:bar_BRAvESP_peak} shows the results of running the appdata algorithm allocating a varying number of extra CPUs when peaks were detected. Just as CPUs allocated by the load algorithm, these CPUs take 60 seconds for being available. The test bed chose for the algorithm was the final match of the Confederations Cup: Brazil vs Spain. That is the most challenging match of the seven, with the most tweets and with the highest peaks and where this algorithm is most necessary.

The appdata algorithm was able to deliver better results already with one extra CPU. Compared to the load algorithm alone, the number of tweet above the SLA dropped from 1.67\% to 1.23\% while the cost increased from 20.97 to 21.27 CPU hours. When more extra CPUs are used, the quality consistently increases while the cost increases. At 10 extra CPUs, only 0.12\% of the tweets miss the SLA but at a considerably higher cost of 34.78 CPU hours. At those points, it means an improvement of 92.81\% with an increase of costs of 63.52\%. When compared to the threshold algorithm, the quality improvement was of 95.24\% with a cost increase of only 12.05\%.

Even if the quality improvement is greater than the cost increase, it is important to note that while the percentage of tweets above the SLA seems to fall linearly, the cost seems to increase exponentially. But since the SLA is very close to being completely met, it is probable that the cost-benefit will still be favorable when this happens.

The current peak detection algorithm has false negatives and that is the reason a number of tweets still miss the SLA. It also has false positives, which results in some CPUs being unnecessarily allocated and, since the algorithm only releases a single CPU at once, excess CPUs can take long to disappear. While the excess CPUs are the reason why costs rise so rapidly in the graph they are also the reason why the number of tweets missing the SLA decreases: excess CPUs can compensate an undetected peak if present at the right time.

%% file: conclusion.tex
\section{Conclusion}

Elasticity is a key feature of cloud computing to meet SLA and budget constraints. This paper introduced a detailed case study of using the data generated by the application itself to trigger auto-scaling operations. We used data from Twitter generated during the FIFA 2013 Confederations Cup and an application that calculates sentiment of users watching the matches. Here are the main lessons from our study.

The load algorithm consistently spends fewer resources than the threshold algorithm and is able to react faster allocating a variable amount of resources at a time. That can only happen because of the knowledge of the delay distribution. Also a basic communication between the application and the PaaS or IaaS level is necessary so the current number of tweets in the system is reported.

The threshold still presents better quality for events with moderate tweet volumes, but its best performance is with a threshold of 60\% CPU usage, way below the most common value of 90\%. For jobs processing fast changing amounts of data, smaller thresholds will behave better. The choice of the parameter of the threshold algorithm must be taken carefully. The value of the threshold has a direct impact on the cost of running the application.

For monitoring events with smaller volumes of data, any algorithm performs well, but the load algorithm consumes fewer resources compared to the other algorithms. For moderate sized events, the threshold algorithm is able to perform slightly better but the load algorithm uses fewer resources. For great events with significant bursts, the appdata algorithm is preferred as it is able to predict peaks and prevent many SLA violations. Though it uses more resources than the load algorithm, it is more likely to meet the SLA. The balance between cost and the necessity of SLA adherence must be considered when choosing the algorithm for such events.

Apart from performance, using application data to trigger auto-scaling operations can open possibilities for service managers to configure their dynamic resource requirements in a different way. Instead of trying to define system level metrics such as CPU or memory consumptions, these service managers can focus more on application characteristics and how they are performing over time.